\documentclass[a4paper,10pt]{scrartcl}
\usepackage{amsmath}
\usepackage{amssymb}
\usepackage{amsfonts}
\usepackage{authblk}
\usepackage{bookmark}
\usepackage[makeroom]{cancel}
\usepackage{enumerate}
\usepackage{etoolbox}
\usepackage[left=2cm,right=2cm,top=2cm,bottom=2cm, nohead]{geometry}
\usepackage[pdftex]{graphicx}
\usepackage{hyperref}
\usepackage{natbib}
\usepackage{pdfpages}
\usepackage{subcaption}
\usepackage{tabularx} 
\usepackage{txfonts}
\usepackage{upgreek}
\usepackage{url}
\usepackage{xcolor}
\usepackage{ragged2e}

\newcommand{\question}[1]{\textbf{\textit{#1}}}

\newcommand{\HRule}{\rule{\linewidth}{0.3mm}}
\newcommand{\dd}{\mathrm{d}}
\renewcommand{\pi}{\uppi}

\renewcommand{\bf}{\textbf}
\newcommand{\ci}{\mathrm{i}}
\renewcommand{\rm}{\textrm}

\DeclareMathOperator{\e}{e}
\newcommand{\enkt}{\ensuremath{\e^{-\frac{h\nu}{kT}}}}
\newcommand{\intzinf}{\ensuremath{\int_0^\infty}}
\newcommand{\emx}{\ensuremath{\e^{-x}}}
\newcommand{\bra}{\langle}
\newcommand{\ket}{\rangle}

\newcommand{\newdocument}[1]{
\clearpage
\addcontentsline{toc}{subsection}{#1}
\setcounter{figure}{0}
\setcounter{table}{0}
\setcounter{footnote}{0}
\setcounter{equation}{0}
}

 \title{\vspace*{-3cm}\line(1,0){482}\\Astronomy from 4 Perspectives: The Dark Universe}
\subtitle{Exercises and Proceedings from the German-Italian WE Heraeus Summer School\\held in 2017 in Heidelberg\\\linethickness{.9mm}\line(1,0){482}}

\author[1]{Stefan S. Brems}
\author[2]{Bj\"orn Malte Sch\"afer}
\author[3]{Niccol\`o Bucciantini}
\author[4]{Hannes Keppler}
\author[5]{Markus P{\"o}ssel}
\author[4]{Jonah Cedric Strau{\ss}}
\author[6]{Matthias Taulien}
\affil[1]{Zentrum f{\"u}r Astronomie der Universit{\"a}t Heidelberg, Landessternwarte, K{\"o}nigstuhl 12, 69117 Heidelberg, Germany \newline e-mail: \texttt{sbrems@lsw.uni-heidelberg.de}}
\affil[2]{Zentrum f{\"u}r Astronomie der Universit{\"a}t Heidelberg, Astronomisches Rechen-Institut, M{\"o}nchhofstra{\ss}e 12-14, 69120 Heidelberg, Germany}
\affil[3]{INAF Osservatorio di Arcetri, Firenze, Italy}
\affil[4]{Universit{\"a}t Heidelberg, Germany}
\affil[5]{Haus der Astronomie and Max Planck Institute for Astronomy, Heidelberg, Germany}
\affil[6]{H\"olderlin-Gymnasium and Hector-Seminar, Heidelberg, Germany}
\date{}

\makeatletter
\patchcmd{\@maketitle}{\LARGE}{\normalsize}{}{}

\def\maketitle{{%
  \renewenvironment{tabular}[2][]
    {\begin{flushleft}}
    {\end{flushleft}}
  \AB@maketitle}}
\makeatother

\begin{document}
\maketitle

\section*{Introduction}
The Heraeus Summer School series "Astronomy from four perspectives"\footnote{\url{http://www.physik.uni-jena.de/didaktik_summerschool.html}}\footnote{\url{http://www.haus-der-astronomie.de/en/events/heraeus-four-perspectives}}, funded by the WE Heraeus Foundation,\footnote{\url{https://www.we-heraeus-stiftung.de/}} draws together teachers and teacher students, astronomers, physicists and astronomy students from Germany and Italy. For each summer school, participants gather at one of the four participating nodes: Heidelberg, Padua, Jena, and Florence. The main goal of the series is to bring astronomy into schools, which is achieved by educating and training the teachers and teacher students.\\

In this e-print, we present the exercises, tutorials, and high-school classroom materials developed during the fifth summer school of the series, which took place at Haus der Astronomie in Heidelberg August 26 -- September 2, 2017.

The {\em tutorials} were prepared beforehand for the participants of the Summer schools, and are suitable for use in teacher training. {\em Classroom materials} were developed mainly during the summer school itself, and are suitable for high-school level teaching. They include question sheets for pupils, and some pointers on where to use the material in the German high school curriculum. 

Both sets of materials address the summer school's four main topics: Supernova cosmology, the virial theorem, rotation curves of galaxies, and the temperature of the cosmic microwave background (CMB).

\section*{Tutorials}
Material for the tutorials for summer school participants are included in these proceedings starting at page \pageref{TutorialTOC}. They consist of exercises, which are in most cases supplemented by worked-out solutions. The exercises and associated (mainly python-)scripts were used to give the attendants of the conference hands-on experience with each of the four cosmology-related topics of the summer schools. Some of these exercises are also suitable for use with advanced pupils. Tutorials include the following resources:
\begin{itemize}
\item \textit{Exercise:} Classical exercise sheets and solutions for the participants
\item \textit{Play with data:} Exercises where participants analyze astronomical data. These require a computer with a working Python 3.x programming environment.\footnote{Python is a popular, freely available programming language. One way of getting it is by installing anaconda: \url{https://www.anaconda.com}}
\item \textit{Script:} The python-3 scripts needed in order to solve the \textit{Play with data exercises}; these are available from the Github repository.
\item \textit{Questions:} More advanced questions that can be tackled using the knowledge gained in the preceding exercises
\end{itemize}

Last but not least, we include some exercises about the home planet of the \textit{little prince} (French original: Le Petit Prince). These exercises are unrelated to the dark universe, but allow students to familiarize themselves with nonlinear gradients in the gravitational potential in an easy-going manner; this idea goes back to Francesco Palla who, until his untimely death in early 2016, was a regular and valued contributor to this summer school series.\footnote{
More information can be found in F. Palla, S. Duvernoy and S. Tobin, {\em The Little Prince's Universe} (NAJS / No art Just Sign, 2017, ISBN 9788894099737)
}

\section*{Accessing additional resources}

All resources for the tutorials (including Python scripts) as well as the ``In the Classroom'' resources developed during the summer schools can be found as a Github repository,
\begin{center}
\url{https://github.com/sbrems/Dark_universe}    
\end{center}
and is free for any non-commercial use. In order to allow users to change and adapt the content to their needs, almost all PDF-files are also accompanied by the *.tex files which were used to compile the sheets. Of course, the authors are not responsible for any changes that might then be applied by a third person.\\
Currently it is also worked on making the material available via astroEDU.\footnote{\url{http://astroedu.iau.org}}

Two of the lectures from the summer school can be found online:
\begin{itemize}
\item \textit{The thermal History of the Universe} by Matthias Bartelmann, video online at\\ \url{https://www.youtube.com/watch?v=m35fXJoQLA0}
\item \textit{Introducting the expanding universe} by Markus P\"ossel, video online at\\
\url{https://www.youtube.com/watch?v=gA-0C-88WbE}\\
and extended lecture notes available at \url{https://arxiv.org/abs/1712.10315}
\end{itemize}

\paragraph{Acknowledgments}
We thank the \textit{Wilhelm and Else Heraeus foundation}\footnote{\url{http://www.we-heraeus-stiftung.de}} for funding the summer schools and thus making these unique conferences possible.

\vfill
\begin{figure}[hp]
\centering
\includegraphics[width=1.\textwidth]{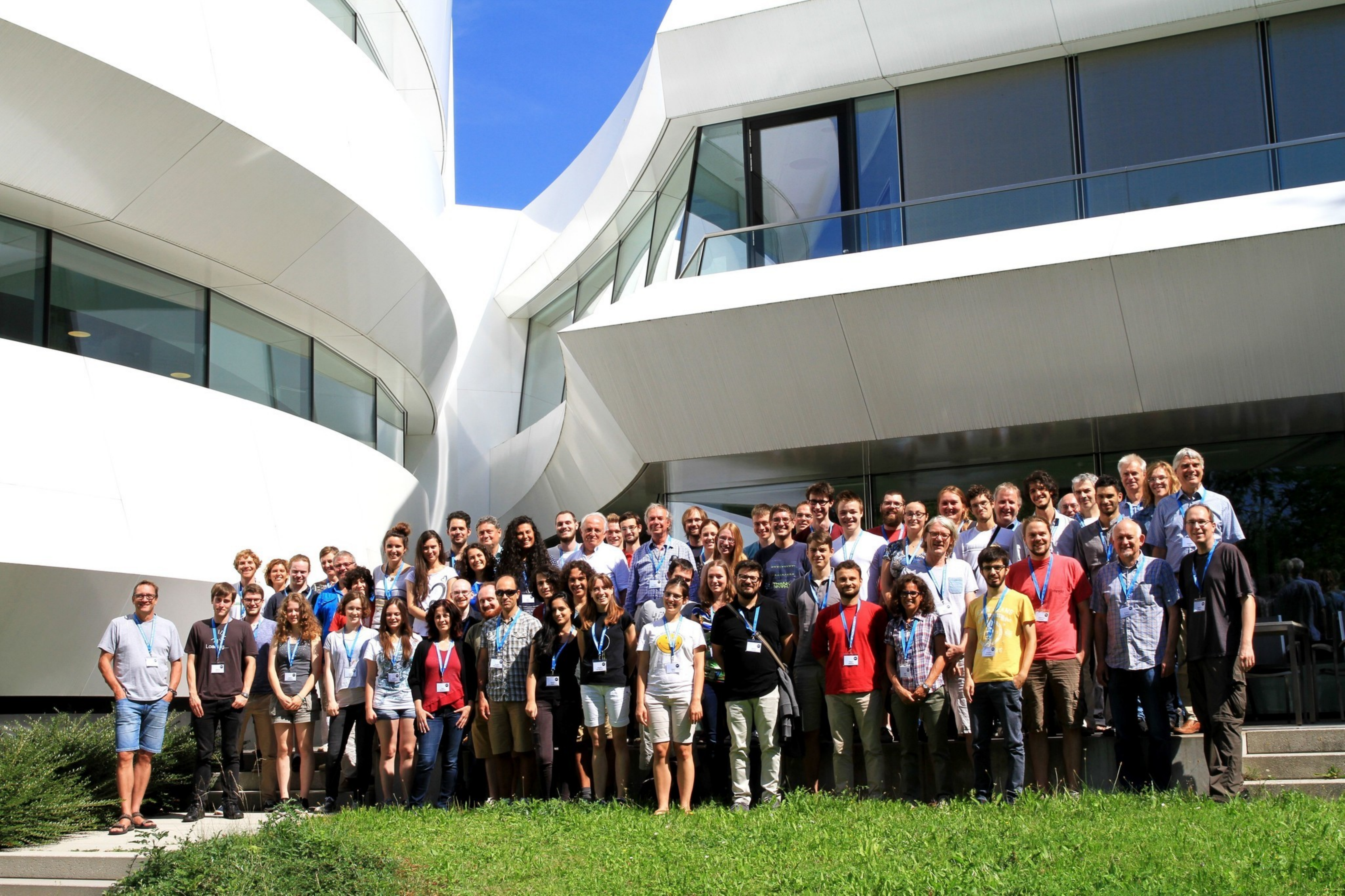}
\caption{Group picture of the summer school 2017 in front of the Haus der Astronomie in Heidelberg.}
\end{figure}
\vfill

\newpage
\centering
\Huge{\textbf{Tutorials: Exercise sheets and solutions}}
\large
\tableofcontents
\newpage
\addcontentsline{toc}{section}{Supernova Cosmology}
\label{TutorialTOC}
\newdocument{Exercises}





\begin{center}
\LARGE \textbf{Astronomy from 4 Perspectives: the Dark Universe}
\HRule
\end{center}
\begin{flushright}
prepared by: Florence participants and Bj{\"o}rn Malte Sch{\"a}fer
\end{flushright}
\begin{center}
{\Large \textbf{Exercise: Supernova cosmology and dark energy}}
\end{center}
\vspace{5mm}

\begin{enumerate}[\itshape \bfseries 1.]

\item \question{Light-propagation in FLRW-spacetimes}\\
Photons travel along null geodesics, $\dd s^2=0$, in any spacetime. 
\begin{enumerate}[(a)]
\item{Please show that by introducing {\em conformal time $\tau$} in a suitable definition, one recovers Minkowski light propagation $c\tau = \pm\chi$ in comoving distance $\chi$ and conformal time $\tau$ for FLRW-space times,
\begin{equation}
\dd s^2 = c^2\dd t^2 - a^2(t)\dd\chi^2,
\end{equation}
which we have assumed to be spatially flat for simplicity.}
\item{What's the relationship between conformal time $\tau$ and cosmic time $t$? What would the watch of a cosmological observer display?}
\item{Please compute the conformal age of the Universe given a Hubble function $H(a)$,
\begin{equation}
H(a) = H_0 a^{-3(1+w)/2}
\end{equation}
which is filled up to the critical density with a fluid with a fixed equation of state $w$.
}
\item{In applying $\dd s^2=0$ to the FLRW-metric we have assumed a radial geodesic - is this a restriction?}
\item{Please draw a diagram of a photon propagating from a distant galaxy to us in conformal coordinates for a cosmology of your choice, with markings on the light-cone corresponding to equidistant $\Delta a$.}
\end{enumerate}

\item \question{Light-propagation in perturbed metrics}\\
The weakly perturbed ($\left|\Phi\right|\ll c^2$) Minkowski metric is given by 
\begin{equation}
\dd s^2 = \left(1+2\frac{\Phi}{c^2}\right) c^2\dd t^2 - \left(1-2\frac{\Phi}{c^2}\right)\dd x_i\dd x^i
\end{equation}
with the Newtonian potential $\Phi$. Please compute the effective speed of propagation $c^\prime = \dd\left|x\right|/\dd t$ for a photon following a null geodesic $\dd s^2=0$. Please Taylor-expand the expression in the weak-field limit $\left|\Phi\right|\ll c^2$: Can you assign an effective index of refraction to a region of space with a nonzero potential?

\item \question{Classical potentials including a cosmological constant}\\
The field equation of classical gravity including a cosmological dark
energy density $\lambda$ is given by
\begin{align}
\Delta\Phi = 4\pi G(\rho + \lambda)
\end{align}
\begin{enumerate}
\item{Solve the field equation for $3$ dimensions outside a
    spherically symmetric and static matter distribution $\rho$. The
    expression for the Laplace-operator in spherical coordinates for
    $3$ dimensions is: $\Delta\Phi=
    r^{-2}\partial_r(r^2\partial_r\Phi)$. Also, please set as the
    total baryon mass $M$
\begin{equation}
M = 4\pi\int_0^r\dd r^\prime (r^\prime)^2\rho(r^\prime)
\end{equation}
}
\item{Please show that both source terms individually give rise to power-law solutions for $\Phi(r)$.}
\item{Is there a distance where the baryon part from the $\rho$-terms
    is equal to dark energy part the $\lambda$-term?}
\item{Assuming a typical galaxy is formed by 100 billion stars like
    the Sun each with a mass of $10^{30}$ kg, and a dark energy density
    of $10^{-27}$ kg / m$^3$, find at which distance from a galaxy the
    dark energy dominates. How does it compare with the typical size
    of a Galaxy ($\sim 10000$ pc?}
\end{enumerate}

\item \question{Physics close to the horizon}\\
Why is it necessary to observe supernovae at the Hubble distance $c/H_0$ to see the dimming in accelerated cosmologies? Please start at considering the curvature scale of the Universe: A convenient measure for the curvature might be the Ricci-scalar $R = 6H^2(1-q)$ for flat FLRW-models.
\begin{enumerate}[(a)]
\item{Can you define a distance scale $d$ or a time scale from $R$?}
\item{What happens on scales $\ll d$, what on scales $\gg d$?}
\end{enumerate}

\item \question{Measuring cosmic acceleration}\\
The luminosity distance $d_\mathrm{lum}(z)$ in a spatially flat FLRW-universe is given by
\begin{equation}
d_\mathrm{lum}(z) = (1+z)\int_0^z\mathrm{d}z^\prime\:\frac{1}{H(z^\prime)}
\end{equation}
with the Hubble function $H(z)$. Let's assume that the Universe is filled with a cosmological fluid up to the critical density with a fluid with equation of state $w$, such that the Hubble function is
\begin{equation}
H(z) = H_0 (1+z)^\frac{3(1+w)}{2}.
\end{equation}
\begin{enumerate}
\item{For this type of cosmology you will obtain acceleration if $w<-1/3$ and deceleration for $w>-1/3$: Please show this by computing the deceleration parameter $q=-\ddot{a}a/\dot{a}^2$ from the Hubble-function $H=\dot{a}/a$, with the relation $a=1/(1+z)$.}
\item{Please show that in accelerated universes supernovae appear systematically dimmer, because $d_\mathrm{lum}$ is always larger than in a non-accelerating universe.}
\item{Is it true that $d_\mathrm{lum}$ is systematically smaller in a decelerating universe?}
\item{Would the expression for $d_\mathrm{lum}$ still be valid if the universe was contracting instead of expanding? What correction would you need to apply?}
\end{enumerate}

\end{enumerate}

\newdocument{Solutions}




	
	\begin{center}
		\LARGE \textbf{Astronomy from 4 Perspectives: the Dark Universe}
		\HRule
	\end{center}
	\begin{flushright}
		prepared by: Florence participants and Bj{\"o}rn Malte Sch{\"a}fer
	\end{flushright}
	\begin{center}
		{\Large \textbf{Exercise: Supernova-cosmology and dark energy}}\\
		\vspace*{2mm}
		{\Large \textbf{Solutions}}
		
	\end{center}
	\vspace{5mm}
	
	\begin{enumerate}[\itshape \bfseries 1.]
		
		\item \question{Light propagation in FLRW-spacetimes}\\
		Photons travel along null geodesics, $\dd s^2=0$, in any spacetime. 
		\begin{enumerate}[(a)]
			\item Let us do the following substitution $dt
                          \rightarrow a(t)d\tau$ then the line element
                          can be written
                          $ds^2=a(t)^2[c^2d\tau^2-d\chi^2]$ and the
                          equation of the null-geodesic will be
                          $d\chi=\pm cd\tau$.
			\item The cosmic time is the time measured by
                          a cosmic observer synchronized for $t=0$ 
 \begin{align}
   t=\int_0^t dt' = \int_0^a\frac{da'}{\dot{a}'}
   \end{align}
The conformal time is tied to the time interval over which an observer
at $t=t_0$ sees to happen an event in the past at time $t$. Now at
$t=t_0$ this will coincide with the cosmic time, hence it will be
affected by cosmic time dilation.
                          \begin{align}
                            \tau(t)=\int_0^t \frac{dt'}{a(t')} = \frac{1}{a(t)}
                              \int_0^t \frac{a(t)}{a(t')} dt' > \frac{t}{a(t)}
                          \end{align}
			\item Now for the given metric:
                          \begin{align}
                            H=\frac{\dot{a}}{a}= H_o a^{-3(1+w)/2}
                            \Rightarrow \dot{a} = H_o
                            a^{1-3(1+w)/2}\\
                            a(t)=\left(\frac{4}{9}\right)^{3(w+1)}(t+wt)^{2/3(w+1)}
                            \end{align}
                          \begin{align}
                            \tau_H=\int_0^t \frac{dt'}{a(t')} =
                            \int_0^1 \frac{da'}{\dot{a}a} =
                            \frac{1}{H_o}\int_0^1 a^{3(w+1)/2 -2}da =  \frac{1}{H_o}\frac{1}{3(w+1)/2 -1}
                            \end{align}
			\item Isotropy of the universe ensures us that
                          it is not.
		\end{enumerate}

		\item \question{Light propagation in perturbed metrics}\\
		\begin{align}
		ds^2=\left(1+2\frac{\Phi}{c^2}\right)c^2dt^2-\left(1-2\frac{\Phi}{c^2}\right)dx^2 
		\end{align}
		With $ds^2=0$:
		\begin{align}
		\left( 1+\frac{2\Phi}{c^2}\right)c^2dt^2 &= \left(1-\frac{2\Phi}{c^2}\right) dx^2\\
		\frac{dx}{dt}&=\pm c\sqrt{\frac{1+\frac{2\Phi}{c^2}}{1-\frac{2\Phi}{c^2}}}
		\end{align}
		With $\frac{1}{1-\epsilon}\approx 1+\epsilon$ for small $\epsilon$:
		\begin{align}
		\frac{dx}{dt}\approx\pm c\left(1+\frac{2\Phi}{c^2}\right)
		\end{align} 
		For a non-zero $\Phi$ this is not equal to $c$! \\
		We assign an effective index of refraction by:
		\begin{align}
		n(\Phi)=\frac{dx/dt}{c}\approx \left(1+\frac{2\Phi}{c^2}\right)
		\end{align}
		
		\item \question{Classical potentials including a cosmological constant}\\
		The field equation of classical gravity including a
                cosmological dark energy density $\lambda$ is given by
		\begin{equation}
		\Delta\Phi = 4\pi G(\rho + \lambda)
		\end{equation}
		(a) field calculation\\
	
		Now it is possible to simply integrate the field equation starting with:
		\begin{align}
		  \Delta\Phi&=\frac{1}{r^{2}}\frac{\partial}{\partial r}\left(r^{2}\frac{\partial\Phi}{\partial r}\right)\\
		  &=4\pi G(\rho(r)+\lambda)\\
		  r^{2}\frac{\partial\Phi}{\partial
                    r}&=\int_0^r\textrm{d}r'\left(4\pi G[\left(r'\right)^{2}\rho\left(r'\right)+\left(r'\right)^{2}\lambda]\right)\\
		  &=GM+G\frac{\lambda}{3}r^3\\
		  \frac{\partial\Phi}{\partial r}&=\frac{GM}{r^{n-1}}+G\frac{\lambda r}{n}\\
		  \Phi&=-\frac{GM}{r}+G\frac{\lambda r^2}{6}
		\end{align}
		(b) power-law solutions\\
		Following the calculation one may see that each source term corresponds to an individual power-law:
		\begin{align*}
		  C(n)G\rho(r) ~~~ &\Rightarrow ~~~ -\frac{GM}{r}\\
		  \lambda ~~~ &\Rightarrow ~~~ G\frac{\lambda r^2}{6}
		\end{align*}
		(c) equilibrium\\
		To find an equilibrium distance one must set \(\Phi\left(r_\textrm{eq}\right)=0\)
		\end{enumerate}
		\begin{align}
		  \frac{GM}{r_\textrm{eq}}&=G\frac{\lambda r_\textrm{eq}^2}{6}\\
		  \frac{\lambda r_\textrm{eq}^3}{6}&=M
		\end{align}
		from which follows immediately:
		\begin{equation}
		  r_\textrm{eq}=\sqrt[3]{6\frac{M}{\lambda}}
		\end{equation}
		(c) if one inputs the number one gets $ r_\textrm{eq}
                =1.5$ Mpc one hundred times larger than the size of a galaxy.
        \begin{enumerate}
        \setcounter{enumi}{3}
		\item \question{Physics close to the horizon}\\
		Why is it necessary to observe supernovae at the Hubble distance $c/H_0$ to see the dimming in accelerated cosmologies? Please start at considering the curvature scale of the Universe: A convenient curvature measure might be the Ricci scalar $R = 6H^2(1-q)$ for flat FLRW-models.
		\begin{enumerate}[(a)]
			\item The Dimension of the Ricci scalar is $1/s^2$ thus we can define a time and a distance scale by:
			$$ \tau = 1/\sqrt{R} \ \ \ \mathrm{and} \ \ d = c/\sqrt{R} \approx c/H_0$$
			which gives the curvature scale of the Universe. 
			\item To observe supernovae dimming caused by accelerated cosmic expansion the supernova distance had to be about (or larger than) the curvature scale, because at distances $<<d$ the different cosmological distance measures converge. \\
			For illustration see: \textit{https://en.wikipedia.org/wiki/Distance\_measures\_(cosmology)}
		\end{enumerate}

		\item \question{Measuring cosmic acceleration}\\
		The luminosity distance $d_\mathrm{lum}(z)$ in a spatially flat FLRW-universe is given by
		\begin{equation}
		d_\mathrm{lum}(z) = (1+z)\int_0^z\mathrm{d}z^\prime\:\frac{1}{H(z^\prime)}
		\end{equation}
		with the Hubble function $H(z)$. Let's assume that the Universe is filled with a cosmological fluid up to the critical density with a fluid with equation of state $w$, such that the Hubble function is
		\begin{equation}
		H(z) = H_0 (1+z)^\frac{3(1+w)}{2}.
		\end{equation}
		\begin{enumerate}
			\item
					By definition:
					\begin{align*}
	H=\frac{\dot{a}}{a} \text{ and } q=-\frac{\ddot{a}a}{\dot{a}^2}
		\end{align*}
		It follows
		\begin{align*}
		\dot{H}&=\frac{\ddot{a}a-\dot{a}^2}{a^2}=\frac{\ddot{a}a}{a^2}-H^2		
		\end{align*}
		So we get
		\begin{align*}
		\frac{\dot{H}}{H^2}&=\frac{\ddot{a}a}{\dot{a}^2}-1=-q-1\\
		q&=-(\frac{\dot{H}}{H^2}+1)	
		\end{align*}

		We also have
		\begin{align*}
		H=H_0\cdot(1+z)^{\frac{3(1+w)}{2}}=H_0\cdot a^{\frac{-3(1+w)}{2}}
		\end{align*}
		and
		\begin{align*}
		\dot{H}&=H_0\left(\frac{-3(1+w)}{2}\right)\cdot a^{\frac{-3(1+w)}{2}}\cdot \dot{a}\\
			  &=H_0\cdot a^{\frac{-3(1+w)}{2}}\cdot\frac{\dot{a}}{a}\cdot\left(\frac{-3(1+w)}{2}\right)\\
			  &=H^2\cdot \left(\frac{-3(1+w)}{2}\right)
		\end{align*}	
		so
		\begin{align*}
		q=-\left(\frac{-3(1+w)}{2}+1\right)=\frac{1}{2}(3w+1)
		\end{align*}
		and obviously
		\begin{align*}
		q<0 \text{ for } w<-\frac{1}{3}\\
		q>0 \text{ for } w>-\frac{1}{3}
		\end{align*}	
										
			\item
				First, we consider the case $w=-\frac{1}{3}$ (non-accelerating universe):
				\begin{align*}
			H=H_0(1+z)^{\frac{3(1+w)}{2}}=H_0(1+z)\\
\end{align*}	
\begin{align*}
d_{lum,1}&=(1+z)\int_0^z \! \frac{1}{H(z')} \, \mathrm{d}z'\\
			&=(1+z)\int_0^z \! \frac{1}{H_0(1+z')} \, \mathrm{d}z'\\
			&=\frac{1+z}{H_0}ln(1+z)
\end{align*}						
	Now, we consider the case $w<-\frac{1}{3}$ (accelerating universe):
	\begin{align*}
				d_{lum,2}&=(1+z)\int_0^z \! \frac{1}{H(z')} \, \mathrm{d}z'\\
				&=\frac{1+z}{H_0}\int_0^z \! (1+z')^{\frac{-3(1+w)}{2}} \, \mathrm{d}z'\\
				&=\frac{1+z}{H_0}\left[(1+z')^{\frac{-3(1+w)+2}{2}}\cdot \frac{2}{-3(1+w)+2}\right]_0^z\\
				&=\frac{1+z}{H_0}\left(\frac{2}{-3(1+w)+2}\right)\left[(1+z')^{\frac{-3(1+w)+2}{2}}-1\right]
\end{align*}	
It follows:
$d_{lum_2}(z)>d_{lum_1}(z)$, because the exponent $\frac{-3(1+w)+2}{2}$ is positive $(w<-\frac{1}{3})$,
so $d_{lum_2}(z)$ is growing faster, than the logarithmic function $d_{lum_1}(z)$.		
			\item Yes, as long as the universe is flat. Just try to plot the two functions
			\item The formula still apply. But in a
                          contracting universe one would see light
                          that is blue-shifted and not red-shifted. Su
                          $z<0$. But obviously $z>-1$, since otherwise one
                          would get negative wavelengths (frequencies)
                          that make no sense. As a consequence, the
                          origin of time in a contracting universe
                          corresponds to $z=-1$ for a universe
                          contracting from infinity. Hence the limit
                          of integration must be corrected accordingly.
		\end{enumerate}
		
	\end{enumerate}
\newdocument{Play with data}






\begin{center}
\LARGE \textbf{Astronomy from 4 Perspectives: the Dark Universe}
\HRule
\end{center}
\begin{flushright}
prepared by: Florence participants and Bj{\"o}rn Malte Sch{\"a}fer
\end{flushright}
\begin{center}
{\Large \textbf{Play with data: Supernova cosmology and dark energy}}
\end{center}
\vspace{5mm}

\noindent
The Supernova Cosmology Project observed supernovae of the type Ia in distant galaxies, determined the distance modulus by measuring the apparent brightness as well as the redshift of the host galaxy. From the relation between distance and redshift one can measure the density parameters $\Omega_X$ and the dark energy equation of state $w$. The exercise sheet uses data from A. Goobar et al., PhST, 85, 47 (2000).

\begin{enumerate}[\itshape \bfseries 1.]

\item \question{Distance-redshift-relationships in FLRW-universes}\\
In this exercise you can play with SCP-data and explore the sensitivity of the supernova brightness on the cosmological parameters. Please have a look at the python-script \path{supernova_plot.py}, which reads the data file from SCP and plots distance modulus $\mu$ as a function of redshift $z$. The cosmological model is a spatially flat FLRW-cosmology with the Hubble function
\begin{equation}
H(z) = H_0\sqrt{\Omega_m(1+z) + (1-\Omega_m)(1+z)^{3(1+w)}}
\end{equation}
where the density of the dark energy component is automatically set to $\Omega_X=1-\Omega_m$ to enforce flatness. Setting $w=-1$ recovers the case of the cosmological constant $\Lambda$, in which case $\Omega_X=\Omega_\Lambda$.

\begin{enumerate}[(a)]
\item{Start by guessing different values for $\Omega_m$ to find the best value: What's $\Omega_\Lambda$?}
\item{What's your explanation why $\Lambda$ makes the supernovae dimmer?}
\end{enumerate}

\begin{figure}[h]
\begin{center}
\resizebox{9cm}{!}{\includegraphics{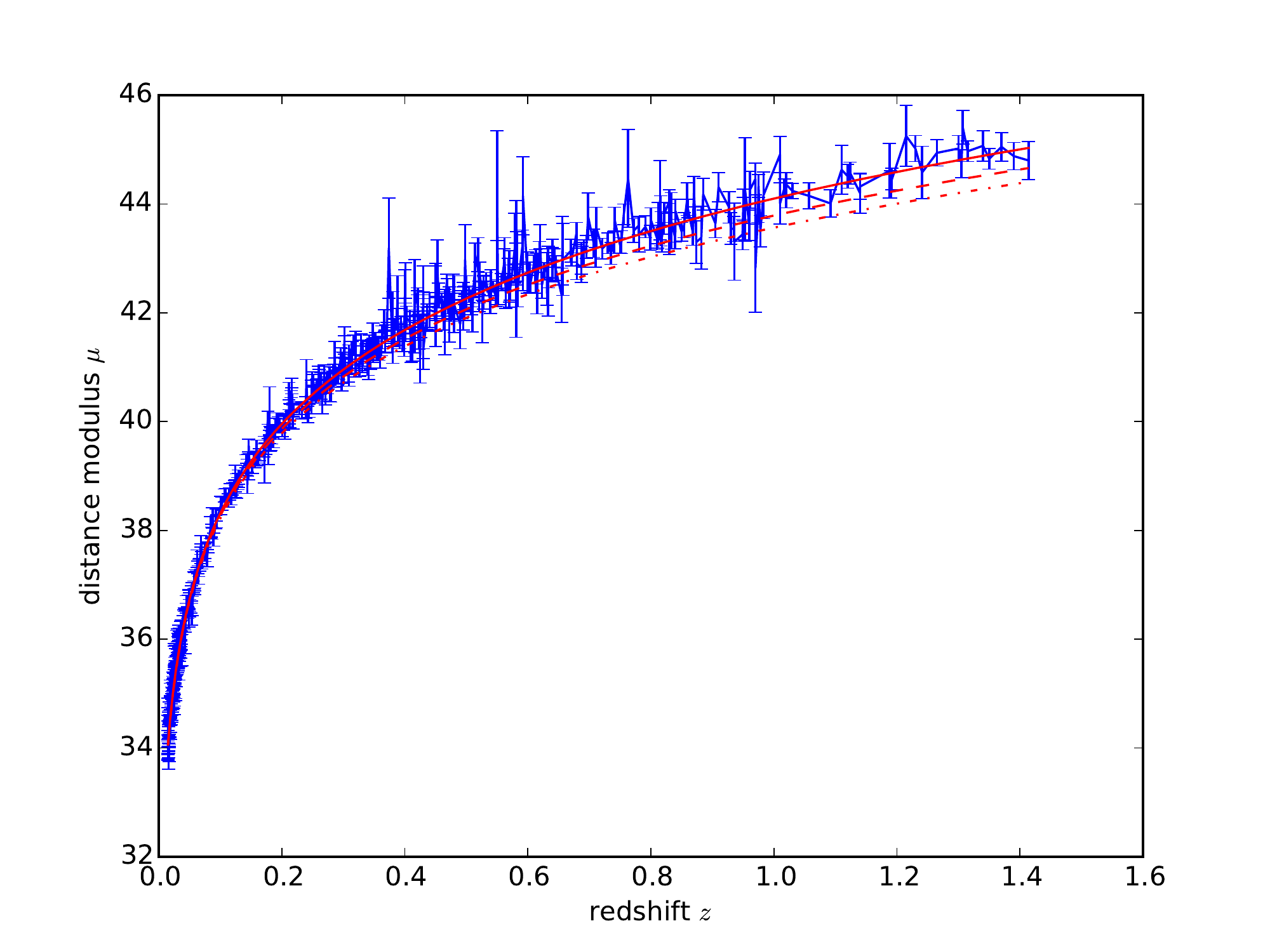}}
\caption{data from the SCP and distance redshift relationships with varying $\Omega_m$}
\end{center}
\end{figure}

\item \question{Fitting a FLRW-cosmology}\\
The script \path{supernova_fit.py} does a proper regression of a model $\mu(z)$ to the data, by minimising the squared difference between data and model, in units of the measurement error. 
\begin{enumerate}[(a)]
\item{What is the best fitting value for $\Omega_m$?}
\item{What's the certainty that $\Omega_\Lambda\neq 0$?}
\end{enumerate}

\begin{figure}[ht]
\begin{center}
\resizebox{9cm}{!}{\includegraphics{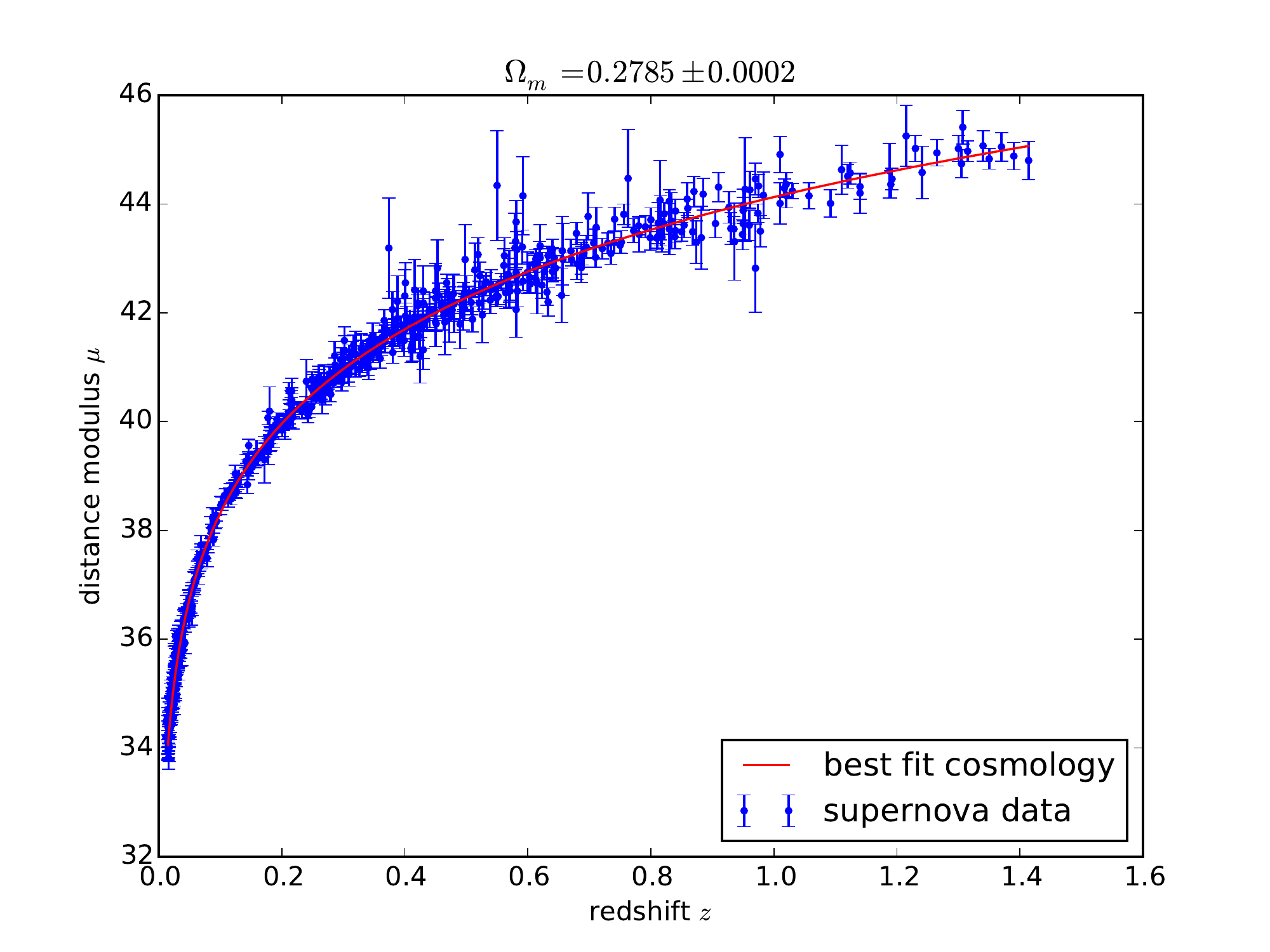}}
\caption{fit to the SCP-data in a FLRW-cosmology: $\Lambda$ is not zero}
\end{center}
\end{figure}

\item \question{Precision of the measurement}\\
In running the script \path{supernova_likelihood.py} you can simultaneously fit $\Omega_m$ and $w$ to the data. It evaluates the likelihood $\mathcal{L}(\Omega_m,w)\propto \exp(-\chi^2(\Omega_m,w)/2)$, with 
\begin{equation}
\chi^2(\Omega_m,w) = \sum_{i=1}^{n_\mathrm{data}}\left(\frac{\mu_i-\mu(z_i,\Omega_m,w)}{\sigma_i}\right)^2
\end{equation}
for the $n_\mathrm{data}$ data points $\mu_i$ at the redshifts $z_i$. The probability that a parameter choice is true is reflected by the density of points.
\begin{enumerate}[(a)]
\item{What are the statistical errors on $\Omega_m$ and on $w$?}
\item{Why do you require more negative $w$ if $\Omega_m$ is larger?}
\end{enumerate}

\begin{figure}[h]
\begin{center}
\resizebox{9cm}{!}{\includegraphics{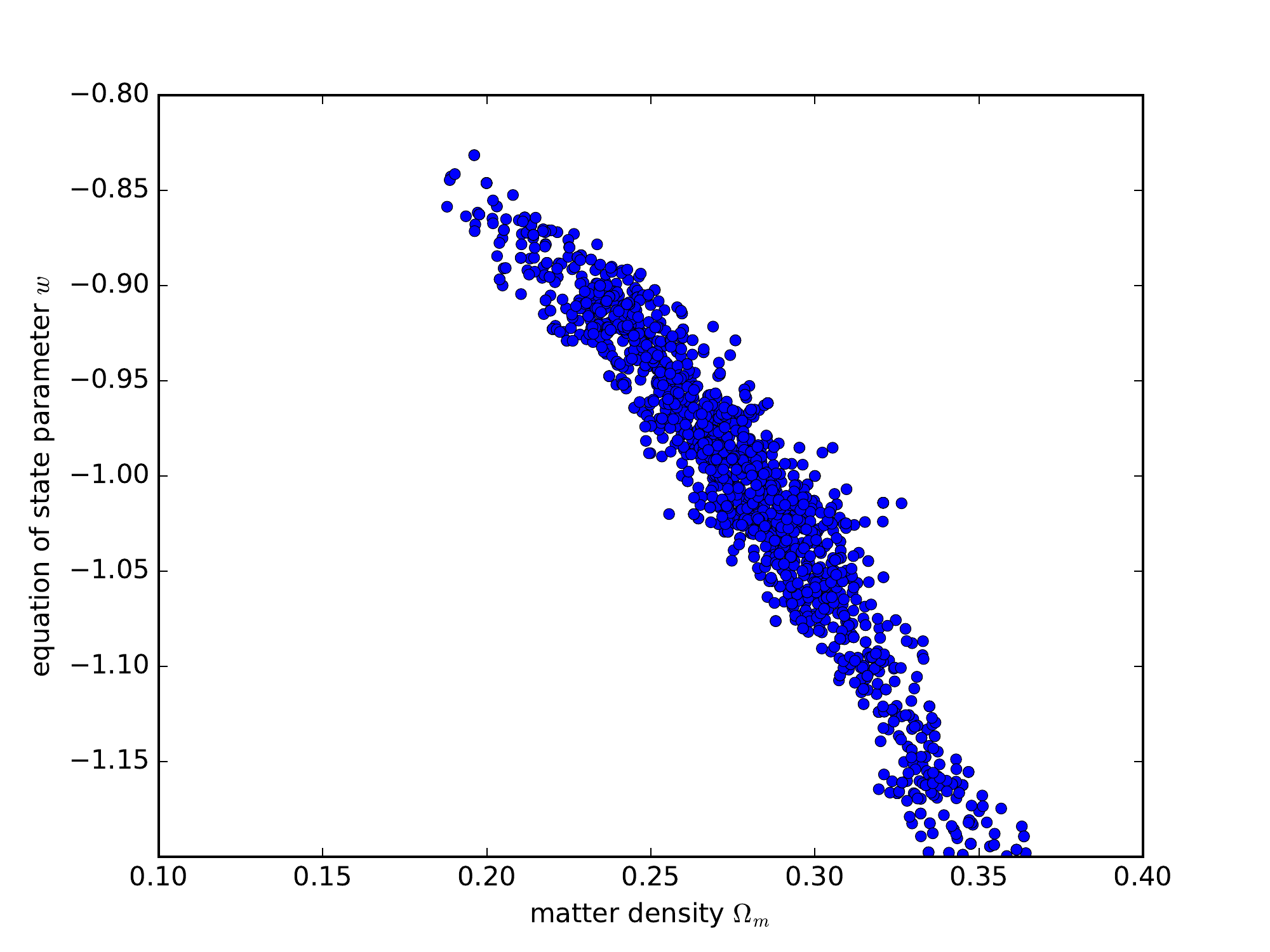}}
\caption{simultaneous measurement of $\Omega_\Lambda$ and $w$}
\end{center}
\end{figure}

\end{enumerate}
\newdocument{Questions}





\begin{center}
\LARGE \textbf{Astronomy from 4 Perspectives: the Dark Universe}
\HRule
\end{center}
\begin{flushright}
prepared by: Florence participants
\end{flushright}
\begin{center}
{\Large \textbf{Questions: Supernova cosmology and dark energy}}
\end{center}
\vspace{5mm}

\begin{enumerate}[\itshape \bfseries 1.]

\item \question{FLRW-models and the equation of state}\\
\begin{enumerate}[(a)]
\item{What are the ingredients (fields?) the enter into a FLRW
    metric?}
\item{How does each field behave as the universe expands?}
\end{enumerate}

\item \question{Light propagation in relativity}\\
\begin{enumerate}[(a)]
\item{What is the equation that describes light propagation in a
    curved space-time?}
\item{How does the presence of a body exerting gravity affect light
    paths?}
\item{How curvature affect light propagation?}
\item{How expansion affects light propagation?}
\end{enumerate}

\item \question{Distance measures}\\
\begin{enumerate}[(a)]
\item{How do we measure distance in a curved space-time?}
\item{What is the angular distance?}
\item{How we disentangle red-shift from curvature?}
\end{enumerate}

\item \question{Supernova cosmology}\\
\begin{enumerate}[(a)]
\item{Why are Supernovae of Type Ia standard candles?}
\item{Why is it important to observe them at high redshift?}
\item{Why are they not enough to fully constrain the cosmology?}
\end{enumerate}

\end{enumerate}
\newdocument{Tutorial: Chandrasekhar mass}





\center
{\LARGE \textbf{Astronomy from 4 Perspectives: the Dark Universe}}
\endcenter

\HRule

\begin{flushright}
prepared by: Niccol{\`o} Bucciantini
\end{flushright}

\center
{\Large \textbf{Quantum Mechanics, Relativity and Supernovae}}
\endcenter

\justify

This tutorial is aimed at advanced students, to whom the teacher has
already introduced the basic concepts of quantum mechanics and
relativity. In particular the only notion of QM required is the {\bf
  uncertainty principle of position and momentum} and the only notion
from SR the {\bf relativistic
relation between energy and momentum}. Some familiarity with
thermodynamics and hydrostatic is also required.

We are going to show how with  simples arguments using basic quantum
mechanics and special relativity one can find that there is a limit to
the mass of a quantum star.\\

The starting points are:
\begin{enumerate}
\item From quantum mechanics the uncertainty principle  of momentum and
position $\Delta x \Delta p
=\hbar$. The teacher should have introduced it before and explained its
meaning and implications.
\item From special relativity the relativistic relation between energy
  and momentum of a relativistic particle $e=cp$. One can use photons
  to show that they carry energy (obvious - think of solar panels that
  produce electricity) and also momentum (the solar mill is a good example).
\item From thermodynamics the relation between pressure of a gas and
  the energy density. This can be introduced classically with a
  discussion of the relation $P = n kT = U$, recalling that $kT$ is
  the thermal energy of a particle of a gas.
\item From hydrostatic the equilibrium relation $F = -\nabla P$,
  stating that any external force (force density to be more precise) must be balanced by a pressure
  gradient. As an example one can discuss how pressure changes going
  under water, by simply showing that at any depth the pressure must
  be equal to the force exerted by the overlying column of water. And
  use this argument to derive the hydrostatic relation. 
\end{enumerate}

We begin by showing how a relativistic gas behaves.\\

First consider a volume V containing N particles. Then the average
volume per particle is $V/N = 1/n$ where we have introduced the
particle density $n$. Then one can define the average distance between particles
is $(V/N)^{1/3} = 1/n^{1/3}$. Using the uncertainty principle, and taking as a
typical uncertainty over the distance the average distance between
particles, we get a typical momentum $p= \hbar n^{1/3}$. Now use the
relativistic energy momentum relation to get a
typical energy $e= c \hbar n^{1/3}$. Finally the energy density of this system of
particles will be $U= n e = c \hbar n^{4/3}$. Recall from classical
thermodynamics that the pressure is of the order of the energy density
then the pressure of our relativistic quantum system
system is $P= c \hbar n^{4/3}$. And setting equal to $m$ the typical
mass of a particle $P=c \hbar (\rho/m)^{4/3} $, where we have
introduced the mas density $\rho$.

We now turn to hydrostatic equilibrium. For a star the equation at
any depth where the density is $\rho$ will look like:
\begin{align}
G\frac{M\rho}{r^2} = -\nabla P = -\frac{dP}{dr}
\end{align}
given that stars are spherically symmetric. At this point we are going
to simplify the treatments looking only at the scaling of the
equation. This is done replacing some of the quantities with their
simplest approximations: the local radius
$r\rightarrow R$ the stellar radius; the density $\rho \rightarrow M
R^{-3}$ where $M$ is the mass of the star; the pressure derivative
$dP/dr \rightarrow P/R$. Then the differential equation turns into an
algebraic one that the students should be more familiar with.
\begin{align}
G\frac{M^2 R^{-3}}{R^2} = -\frac{P}{R}\\
G\frac{M^2}{R^5}  =\frac{ \hbar c}{R} \left(\frac{M}{mR^3}\right)^{4/3}
\end{align}
The student should see immediately that the radius simplifies out of
the equation. This means that the equilibrium is independent of the
radius. Or stated in other words that if the equilibrium (the above
equation) is not satisfied, the star will start to expand (explode) or
collapse (implode), and you cannot avoid this by adjusting the radius;
it is a catastrophic process. The student should also see that the
equilibrium equation gives the following solution of the mass:
\begin{align}
M_{\rm eq} = \frac{1}{m^2}\left(\frac{c \hbar}{G}   \right)^{3/2}
\end{align}
Now let us put the numbers: the speed of light $c=3\times 10^9$ m
s$^{-1}$; $m$ is the mass of a proton $1.6 \times
10^{-27}$ kg; $G = 6.6 \times 10^{-11}$ m$^3$ s$^{-2}$ kg$^{-1}$;
$\hbar=10^{-34}$ J s. Substituting these values one gets:
\begin{align}
M_{\rm eq} = 3.8 \times 10^{30} {\rm kg}
\end{align}
This is about 2 times that mass of the Sun $M_{\rm Sun} = 2\times
10^{30}$ kg (doing the correct model the astrophysicists get 1.4 times
the mass of the Sun - be happy with such simple equation you got only
a 25\% difference). This equilibrium mass is known as {\bf Chandrasekhar mass}. If
the mass is smaller, the star will expand until the physical processes
of the gas change so much that somehow the star reaches a new
equilibrium. If the stellar mass gets bigger than the gravity wins and the star will
start to collapse, driving a catastrophic evolution, that can either
end with a black hole or with a stellar detonation, known as Supernova.

\vspace{5mm}


\addcontentsline{toc}{section}{Dark Matter and the Virial Theorem}
\newdocument{Exercises}





\begin{center}
\LARGE \textbf{Astronomy from 4 Perspectives: the Dark Universe}
\HRule
\end{center}
\begin{flushright}
prepared by: Heidelberg participants
\end{flushright}
\begin{center}
{\Large \textbf{Exercise: Dark matter and the virial theorem}}
\end{center}
\vspace{5mm}

\begin{enumerate}[\itshape \bfseries 1.]

\item \question{Empirical approach to the virial theorem}\\
Please complete this table and compute the specific kinetic energy $T$, the specific potential energy $V$ and the ratio between the two. Does the virial law hold as well for specific kinetic and potential energies? You find the necessary data on all planets on Wikipedia, and please assume that the planets follow circular orbits.

\begin{table}[h]
\begin{center}
\begin{tabular}{|l|ll|ll|l|}
\hline
planet & distance $r$ & orbital period $t$ & kinetic energy $T$ & potential energy $V$ & ratio $T/V$\\
\hline
Mercury & & & & & \\
Venus & & & & & \\
Earth & & & & & \\
Mars & & & & & \\
Jupiter & & & & & \\
Saturn & & & & & \\
Uranus & & & & & \\
Neptune & & & & & \\
\hline
\end{tabular}
\end{center}
\end{table}

\item \question{Kepler orbits and the virial theorem}\\
Why do the planets follow orbits with a fixed ratio between kinetic and potential energy?
\begin{enumerate}[(a)]
\item{Please start by deriving a relationship between orbital velocity $\upsilon$ and distance from a Newtonian calculation for a circular orbit.}
\item{For that orbit, predict the kinetic energy from the potential energy. Is there a fixed ratio between the two?}
\item{The virial theorem is only valid for time-averaged quantities: Why are the energies constant for a circular orbit, implying that you don't have to average?}
\end{enumerate}

\item \question{Relationship to flat rotation curves}\\
An important model for the distribution of (dark) matter inside a halo is the density profile $\rho\propto r^{-2}$ (called isothermal sphere), which has a number of important consequences:
\begin{enumerate}[(a)]
\item{Please compute the potential $\Phi$ of a density profile $\rho\propto r^{-2}$. This type of density profile is typical for dark matter halos at intermediate distances. The solution for $\Phi$ follows from the Poisson equation $\Delta\Phi = 4\pi G\rho$ assuming spherical symmetry,
\begin{equation}
\Delta\Phi = \frac{1}{r^2}\frac{\dd}{\dd r}\left(r^2\frac{\dd\Phi}{\dd r}\right) = 4\pi G\rho.
\end{equation}
}
\item{Now, derive a relationship between the velocity $\upsilon(r)$ of a star following a circular orbit at the distance $r$ from the halo centre: Do you find that $\rho\propto r^{-2}$ enforces $\upsilon(r) = \mathrm{const}$?}
\item{Next, solve the equation of motion $\ddot{r} = -\nabla\Phi$ for a star oscillating in that halo through the centre. Do you find a consequence of the specific profile $\rho\propto r^{-2}$?}
\item{What's the escape velocity from a singular isothermal sphere?}
\end{enumerate}

\item \question{Virial theorem for the harmonic oscillator}\\
Please show for a harmonic pendulum $\ddot{x} = -g/l\:x$ (with the gravitational acceleration $g$ and the pendulum length $l$) that the
\begin{enumerate}[(a)]
\item total energy is conserved at every instant $t$.
\item average kinetic and potential energies are equal. Please use
\begin{equation}
\bra x^2\ket = \frac{1}{\tau}\int_0^\tau\dd t\:x^2(t)
\quad\mathrm{and}\quad
\bra \dot{x}^2\ket = \frac{1}{\tau}\int_0^\tau\dd t\:\dot{x}^2(t)
\end{equation}
as definitions of the average, with the oscillation period $\tau = 2\pi\sqrt{l/g}$.
\end{enumerate}

\item \question{Mechanical similarity and the virial theorem}\\
Mechanical similarity implies the relationship $r^{2-n}\propto t^2$ between the length scale $r$ and the time scale $t$ in mechanical systems with a potential $\Phi\propto r^n$. Collecting results for the 4 most common potentials leads to:

\begin{table}[h]
\begin{center}
\begin{tabular}{|l|lll|}
\hline
system & potential & similarity & remark\\
\hline
Kepler-problem & $\Phi\propto r^{-1}$ & $r^3\propto t^2$ & Kepler's law\\
flat potential & $\Phi=\mathrm{const}$ & $r\propto t$ & inertial motion\\
inclined plane & $\Phi\propto r$ & $r\propto t^2$ & constant acceleration\\
pendulum & $\Phi\propto r^2$ & $t = \mathrm{const}$ & isochrony\\
\hline
\end{tabular}
\end{center}
\end{table}

\begin{enumerate}[(a)]
\item{Why can the virial theorem only be applied to the first and the last case?}
\item{Can you guess with you knowledge of the Kepler law that kinetic and potential energy need to be proportional to each other?}
\item{Boosting into another frame by doing a Galilei transform changes the kinetic energy: Would this affect the virial theorem?}
\end{enumerate}

\item \question{Application to galaxy clusters}\\
The galaxies inside a cluster have kinetic energies that are a factor of $\sim100$ too large, if only the visible matter gravitates: Could you reconcile this by changing the gravitational potential from $\Phi\propto 1/r$ to $\Phi\propto 1/r^n$? Can you predict a number for $n$ from the virial theorem?

\end{enumerate}

\newdocument{Solutions}





	
	\begin{center}
		\LARGE \textbf{Astronomy from 4 Perspectives: the Dark Universe}
		\HRule
	\end{center}
	\begin{flushright}
		prepared by: Heidelberg participants
	\end{flushright}
	\begin{center}
		{\Large \textbf{Exercise: Dark matter and the virial theorem}}\\
		\vspace*{2mm}
		{\Large \textbf{Solutions}}
		
	\end{center}
	\vspace{5mm}
	
	\begin{enumerate}[\itshape \bfseries 1.]
		
		\item \question{Empirical approach to the virial theorem}\\
		Please complete this table and compute the specific kinetic energy $T$, the specific potential energy $V$ and the ratio between the two. Does the virial law hold as well for specific kinetic and potential energies? You find the necessary data on all planets on Wikipedia, and please assume that the planets follow circular orbits.
		
		The specific kinetic energy i. e. kinetic energy divided by mass can be obtained by
		\begin{equation}
		T = \frac{1}{2}\left(\frac{2\pi r}{t}\right)^{2}
		\end{equation}
		
		and the specific potential energy i. e. potential energy divided by mass by
		\begin{equation}
		V = -\frac{GM}{r}
		\end{equation}
		
		where $G$ is Newton's gravitational constant and $M$ the mass of the sun.
		
		\begin{table}[h]
			\begin{center}
            \bgroup
            \renewcommand{\arraystretch}{1.2}
				\begin{tabular}{|c|cc|cc|c|}
					\hline
					planet & distance $r$ & orbital period $t$ & kinetic energy $T$ & potential energy $V$ & ratio $T/V$\\
					& $10^{9}$ m & days & J/kg & J/kg & \\
					\hline
					Mercury & 58 & 88 & $1,2\cdot 10^{9}$ & $-2,3\cdot 10^{9}$ & -0,50\\
					Venus & 108 & 225 & $6,1\cdot 10^{8}$ & $-1,2\cdot 10^{9}$ & -0,49\\
					Earth & 150 & 365 & $4,5\cdot 10^{8}$ & $-8,9\cdot 10^{8}$ & -0,50\\
					Mars & 228 & 687 & $2,9\cdot 10^{8}$ & $-5,9\cdot 10^{8}$ & -0,50 \\
					Jupiter & 778 & 4330 & $8,5\cdot 10^{7}$ & $-1,7\cdot 10^{8}$ & -0,50 \\
					Saturn & 1434 & 10585 & $4,8\cdot 10^{7}$ & $-9,3\cdot 10^{7}$ & -0,52 \\
					Uranus & 2872 & 30660 & $2,3\cdot 10^{7}$ & $-4,6\cdot 10^{7}$ & -0,50 \\
					Neptune & 4498 & 60225 & $1,5\cdot 10^{7}$ & $-3,0\cdot 10^{7}$ & -0,50 \\
					\hline
				\end{tabular}
                \egroup
			\end{center}
		\end{table}
		
		\setcounter{equation}{0}
		\item \question{Kepler orbits and the virial theorem}
		\begin{enumerate}[(a)]
			\item
			
			The gravitational force $F_G$ acts as centripetal force $F_c$:
			\begin{align*}
			F_c &=F_G\\
			\frac{mv^2}{r}&=G\frac{mM}{r^2}
			\end{align*}
			with $r$ as radius of the circle, $v$ as velocity, $m$ as mass of the planet and $M$ as mass of the sun.\\
			It follows
			\begin{align*}
			v^2=\frac{GM}{r}
			\end{align*}
			
			\item
			
			The kinetic energy is $T=\frac{1}{2}mv^2$. So if we use the formula of the potential energy 
			\begin{align*}
			V=-G\frac{mM}{r}
			\end{align*}
			and the result a), it follows
			\begin{align*}
			T=\frac{1}{2}mv^2=\frac{1}{2}m\cdot\frac{GM}{r}=-\frac{1}{2}V
			\end{align*}
			\item
			
			For a circular orbit, the radius $r$ is constant, thus also the potential energy $V$.\\
			The total energy $E=T+V$ is also a constant.\\
			It follows that $T$ is constant.
		\end{enumerate}
		
		\setcounter{equation}{0}
		\item \question{Relationship to flat rotation curves}
		\begin{enumerate}[(a)]
			\item The Poisson equation reads: \\
			
			\begin{align*}
			\nabla  \Phi = \frac{1}{r^2} \frac{d}{dr} (r^2 \frac{d\Phi}{dr}) = 4 \pi G \rho
			\end{align*}
			
			For the density of a SIS profile, we have $\rho \propto r^{-2}$ and hence $\rho = \rho_0 \cdot \frac{r_0}{r}^2$.\\             
			This yields: \\
			\begin{align*}
			\frac{d}{dr} (r^2 \frac{d\Phi}{dr}) &= 4 \pi G \rho_0 r_0^2 \\
			r^2 \frac{d\Phi}{dr} &= \int 4 \pi G \rho_0 dr = 4 \pi G \rho_0 r_0^2  r \\
			\frac{d\Phi}{dr} &= 4 \pi G \rho_0 r_0^2 r^{-1} \\
			\Phi(r) &= \int 4 \pi G \rho_0 r_0^2 r^{-1} dr \\
			\end{align*}
			
			This leads to our final result: \\
			\begin{align*}
			\Phi(r) = 4 \pi G \rho_0 r_0^2 \ln \frac{r}{r_0}
			\end{align*}

			\item The orbital velocity is obtained by equating gravitational and centripetal force: 
			
			\begin{align*}
			F_c &= - F_g \\
			m \frac{v(r)^2}{r} = m \nabla \Phi &= m \frac{d}{dr} 4 \pi G \rho_0r_0^2 \ln \frac{r}{r_0}
			\end{align*}
			
			This yields: 
			\begin{align*}
			m \frac{v(r)^2}{r} &= m 4 \pi G \rho_0 r_0^2 \frac{1}{r}\\
			v(r) = \sqrt{4 \pi G \rho_0 r_0^2} \\
			\end{align*}
			
			This result is independent of $r$; hence, the orbital velocity for objects in a SIS halo is constant even 				at large radii.
			
			\item The equation of motion is given by:
			\begin{align*}
			\ddot{r} &= - \nabla \Phi \\
			\end{align*}

			Setting $k = 4 \pi G \rho_0 r_0^2$, we obtain:
			
			\begin{align*}
			\ddot{r} &= - \nabla \Phi \\
			\ddot{r} &= - \frac{\partial}{\partial r} k \ln \frac{r}{r_0} \\
			\ddot{r} &= - \frac{k}{r} \\
			\end{align*}

			This differential equation is solved by the following expression, where $erf(x) = \frac{1}{\sqrt{\pi}} \int_{-x}^{x} e^{-t^2}dt$ denotes the error function: 
			
			\begin{align*}
			r(t) &= \exp\left[\frac{c_0 - 2 \ erf^{-1}\left(-\sqrt{\frac{2}{\pi}} \sqrt{k e^{-\frac{c_0}{k}}(c_1 + t)^2}\right)^2}{2 k}\right]
			\end{align*}
			
			\item The escape velocity can be obtained by equating the kinetic energy of the escaping body with the work required to move the body from $r_0$ to infinity against the gravitational force: 
			\begin{align*}
			E_{kin} = \frac{m}{2} v_{esc}^2 = W &= - \int_{r_0}^{\infty} F_g \ dr = m \int_{r_0}^{\infty} \nabla \Phi  r \ dr = m \int_{r_0}^{\infty} \nabla \Phi dr 
			\end{align*}
			
			However, this turns out to be infinite: 
			
			\begin{align*}
			W &= 4 \pi G \rho_0 r_0^2 m \int_{r_0}^{\infty} \frac{1}{r} dr
			\end{align*}
			
			Hence, the escape velocity from a SIS halo is infinite. This is due to the unphysical density singularity at the halo centre, which means that the total mass of the halo is also divergent. 
			
		\end{enumerate}
		
		\newpage
		
		\setcounter{equation}{0}
		\item \question{Virial theorem for the harmonic oscillator}
		
		The general solution for the harmonic oscillator
		\begin{equation}
		\ddot{x}=-\omega^2x
		\end{equation}
		with \(\omega^2=g/l\) and the initial conditions (\(x(0)=x_0, \dot{x}(0)=v_0\)) is:
		\begin{equation}
		x(t)=x_0\cos\omega t+\frac{v_0}{\omega}\sin\omega t
		\end{equation}
		
		\begin{enumerate}[(a)]
			\item Energy conservation
		\end{enumerate}
		
		The kinetic energy is:
		\begin{align}
		T&=\frac{1}{2}\dot{x}^2\\
		&=\frac{1}{2}\left(v_0\cos\omega t-x_0\omega\sin\omega t\right)^2\\
		&=\frac{1}{2}\left(v_0^2\cos^2\omega t-2x_0v_0\omega\sin\omega t\cos\omega t+x_0^2\omega^2\sin^2\omega t\right)\\
		&=\frac{v_0^2+x_0^2\omega^2}{4}+\frac{v_0^2-x_0^2\omega^2}{4}\cos 2\omega t-\frac{x_0v_0\omega}{2}\sin2\omega t
		\end{align}
		Whereas the potential energy satisfies:
		\begin{align}
		V&=\frac{1}{2}\omega^2x^2\\
		&=\frac{1}{2}\omega^2\left(x_0\cos\omega t+\frac{v_0}{\omega}\sin\omega t\right)^2\\
		&=\frac{1}{2}\left(\omega^2x_0^2\cos^2\omega t+2x_0v_0\omega\sin\omega t\cos\omega t+v_0^2\sin^2\omega t\right)\\
		&=\frac{x_0^2\omega^2+v_0^2}{4}+\frac{x_0^2\omega^2-v_0^2}{4}\cos 2\omega t+\frac{x_0v_0\omega}{2}\sin2\omega t
		\end{align}
		from which follows, that:
		\begin{align}
		E&=T+V\\
		&=\frac{v_0^2+x_0^2\omega^2}{2}
		\end{align}
		which is constant for all times \(t\) because it is time independent.
		
		\begin{enumerate}[(b)]
			\item Energy equalities
		\end{enumerate}
		
		To calculate the average, one may use the identities obtained above using \(\tau=2\pi/\omega\)
		\begin{align}
		\left\langle x^2\right\rangle&=\frac{1}{\tau}\int_0^\tau\mathrm{d}t~x^2(t)\\
		&=\frac{1}{\tau}\int_0^\tau\mathrm{d}t\left(\frac{x_0^2+v_0^2/\omega^2}{2}+\frac{x_0^2-v_0^2/\omega^2}{2}\cos 2\omega t+\frac{x_0v_0}{\omega}\sin2\omega t\right)\\
		&=\frac{\omega}{2\pi}\left[\frac{x_0^2+v_0^2/\omega^2}{2}t+\frac{x_0^2-v_0^2/\omega^2}{4\omega}\sin 2\omega t-\frac{x_0v_0}{2\omega^2}\cos2\omega t\right]^{2\pi/\omega}_0\\
		&=\frac{\omega}{2\pi}\left[\frac{x_0^2+v_0^2/\omega^2}{2}\frac{2\pi}{\omega}\right]\\
		&=\frac{x_0^2+v_0^2/\omega^2}{2}
		\end{align}
		And for the kinetic terms:
		\begin{align}
		\left\langle\dot{x}^2\right\rangle&=\frac{1}{\tau}\int_0^\tau\mathrm{d}t~\dot{x}^2(t)\\
		&=\frac{1}{\tau}\int_0^\tau\mathrm{d}t\left(\frac{v_0^2+x_0^2\omega^2}{2}+\frac{v_0^2-x_0^2\omega^2}{2}\cos 2\omega t-x_0v_0\omega\sin2\omega t\right)\\
		&=\frac{\omega}{2\pi}\left[\frac{v_0^2+x_0^2\omega^2}{2}t+\frac{v_0^2-x_0^2\omega^2}{4\omega}\sin 2\omega t-\frac{x_0v_0}{2}\cos2\omega t\right]^{2\pi/\omega}_0\\
		&=\frac{\omega}{2\pi}\left[\frac{v_0^2+x_0^2\omega^2}{2}\frac{2\pi}{\omega}\right]\\
		&=\frac{v_0^2+x_0^2\omega^2}{2}
		\end{align}
		So one may see that with:
		\begin{align}
		\left\langle T\right\rangle&=\frac{1}{2}\left\langle\dot{x}^2\right\rangle\\
		&=\frac{v_0^2+x_0^2\omega^2}{4}\\
		\left\langle V\right\rangle&=\frac{1}{2}\omega^2\left\langle x^2\right\rangle\\
		&=\frac{x_0^2\omega^2+v_0^2}{4}
		\end{align}
		The virial theorem \(\left\langle T\right\rangle=\left\langle V\right\rangle\) holds stand.
		
		\setcounter{equation}{0}
		\item \question{Mechanical similarity and the virial theorem}
		\begin{enumerate}[(a)]
			\item Why can the virial theorem only be applied to the first and last case? \\
			
			For the virial theorem to be applied the motion must be constraint in space and momentum, because of the averaging. (see below for more information)  \\
			
			\mbox{ \centering
				\renewcommand{\arraystretch}{1.2}
				\begin{tabular}{|r|l|}
					\hline
					Kepler &  ellipse $\Rightarrow \exists ~ r_{max}, p_{max} < \infty$ \\ \hline
					flat potential &  $r \xrightarrow{t \rightarrow \infty} \infty$, if $p \neq 0$ \\ \hline
					inclined plane	& $r, p \xrightarrow{t \rightarrow \infty} \infty$ \\ \hline
					pendulum & $\exists ~ \theta_{max}, p_{max} < \infty$ \\ \hline
				\end{tabular}}
				\vspace{0.5cm} \\
				\textit{Derivation of the virial theorem:} \\
				
				\begin{align}
				&2T = m\dot{x}^2 = p\dot{x} = \frac{\mathrm{d}}{\mathrm{d}t} (px) - \dot{p}x = \frac{\mathrm{d}}{\mathrm{d}t} (px) + x \frac{\partial V}{\partial x} \\
				&\Rightarrow 2T - x \frac{\partial V}{\partial x} = \frac{\mathrm{d}}{\mathrm{d}t} (px) \\
				&\Rightarrow \langle 2T - x \frac{\partial V}{\partial x}\rangle = \langle \frac{\mathrm{d}}{\mathrm{d}t} (px) \rangle = \lim\limits_{t \rightarrow \infty} \frac{1}{t} \int_{0}^{t} \mathrm{d}t' \frac{\mathrm{d}}{\mathrm{d}t'} (px) \\
				& = \lim\limits_{t \rightarrow \infty} \frac{1}{t} \left[ p(t)x(t)-p(0)x(0) \right] = 0  \ \text{(If p and x are constrained)}\\
				&\Rightarrow 2 \langle T \rangle = \langle x \frac{\partial V}{\partial x} \rangle\\
				&\Rightarrow 2 \langle T \rangle = k \langle V \rangle \ \text{(If V is homogeneous of grade k)}
				\end{align} 
				\item Can you guess with you knowledge of the Kepler law that kinetic and potential energy need to be proportional to each other?\\
				From mechanical similarity follows $r^3 \propto t^2 \Rightarrow (t/r)^2 \propto r^{-1}$ \\
				\item Boosting into another frame by doing a Galilei-transform changes the kinetic energy: Would this affect the virial theorem?\\
				Yes. For an observer moving relative to the system, the motion is in general not constraint.
				$$ T' = T + T_{boost} \Rightarrow \langle T' \rangle = \langle T \rangle + T_{boost}$$
			\end{enumerate}
			
			\setcounter{equation}{0}
			\item \question{Application to galaxy clusters}\\
			For $\Phi\propto r^{-1}$ the virial theorem is 
			\begin{align}
			\left< T\right> &= -\frac{1}{2}\left< V\right>\\
			&\propto -\frac{1}{2}r^{-1}\\
			\end{align}
			If we measure kinetic energies too large by a foctor $\approx 100$ we get $\left < T'\right >=100\left < T\right >$ and can explain this by changing the potential to $\Phi\propto r^{-n}$:
			\begin{align}
			\left < T'\right > &= -\frac{n}{2}\left < V'\right >\\
			100\left < T \right > &\propto -\frac{n}{2} r^{-n}
			\end{align} 
			Because this must hold for all $r$ we predict $n=100$ to explain the measured kinetic energies.

		\end{enumerate}
\newdocument{Play with data}






\begin{center}
\LARGE \textbf{Astronomy from 4 Perspectives: the Dark Universe}
\HRule
\end{center}
\begin{flushright}
prepared by: Heidelberg participants and Bj{\"o}rn Malte Sch{\"a}fer
\end{flushright}
\begin{center}
{\Large \textbf{Play with data: virial theorem and periodic motion}}
\end{center}
\vspace{5mm}

\noindent
In these exercises, we will explore the virial theorem by solving equations of motions numerically and measuring averaged energies from the solutions.

\begin{enumerate}[\itshape \bfseries 1.]

\item \question{Virial theorem and the harmonic oscillator}\\
The script \path{harmonic.py} generates a solution to the harmonic oscillator equation $\ddot{x} = -x$ by transforming it with the definition $y=\dot{x}$ into a coupled first order equation,
\begin{equation}
\frac{\dd}{\dd t}
\left(
\begin{array}{c}
x\\
y
\end{array}
\right)
=
\left(
\begin{array}{cc}
0 & 1\\
-1 & 0
\end{array}
\right)
\left(
\begin{array}{c}
x\\
y
\end{array}
\right),
\end{equation}
where the angular frequency is set to $\omega=1$ for the numerics.
\begin{enumerate}[(a)]
\item{Is the total energy conserved?}
\item{Please run the script and measure the average kinetic and potential energies: Do you find $\bra T\ket = \bra V\ket$ for the harmonic oscillator?}
\item{Is $\bra T\ket = \bra V\ket$ true for any initial condition?}
\end{enumerate}

\begin{figure}[h]
\begin{center}
\resizebox{7.5cm}{!}{\includegraphics{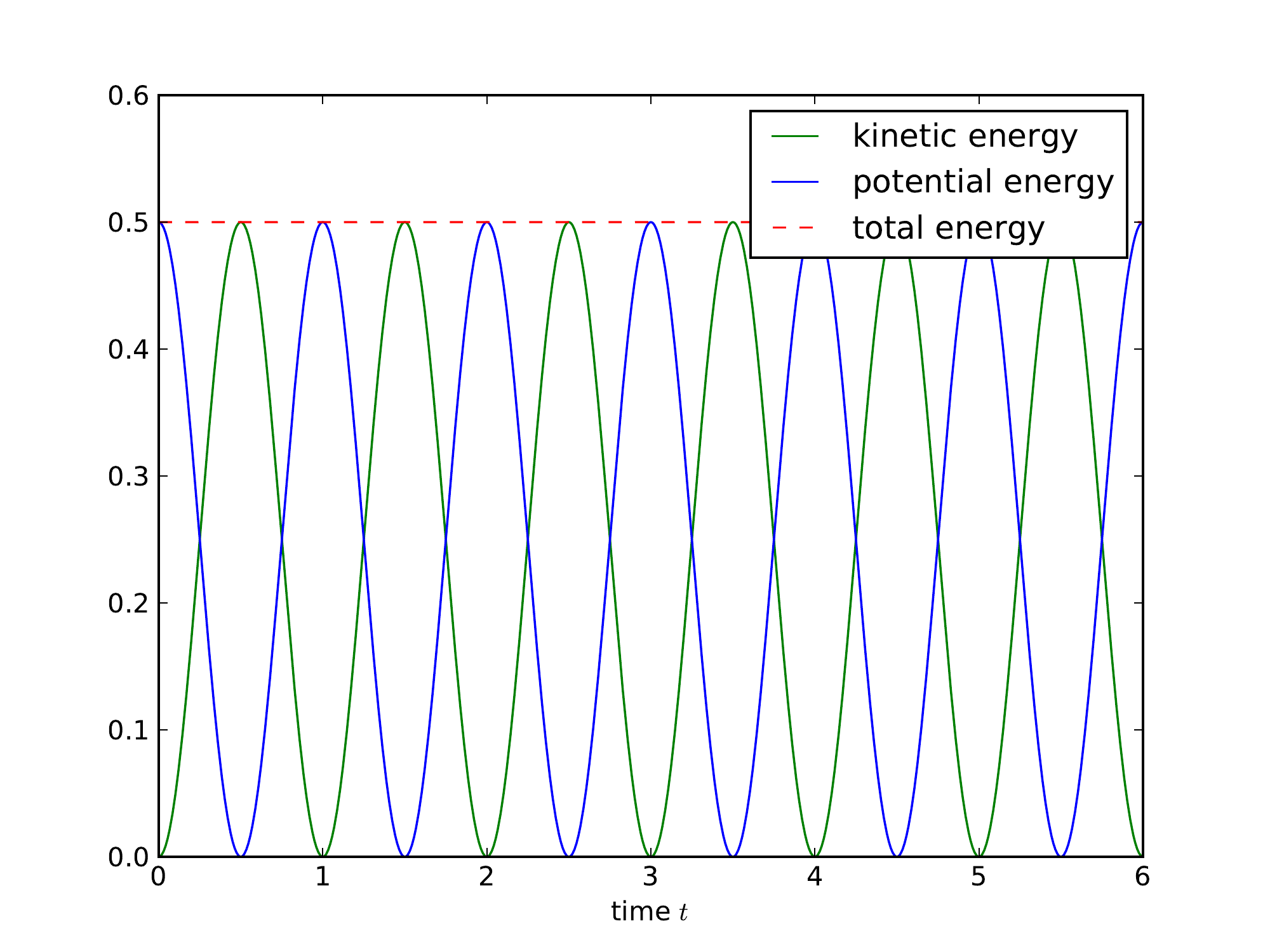}}
\caption{numerical solutions to the harmonic oscillator}
\end{center}
\end{figure}

\item \question{Virial relationship in the anharmonic oscillator}\\
The virial theorem makes a prediction for the average kinetic and potential energies in any system with a scale-free potential, for instance the anharmonic oscillator with the potential $\Phi\propto x^{2n}$. The script \path{anharmonic.py} solves the equation of motion. The constant of proportionality is set to 1.
\begin{enumerate}[(a)]
\item{What's the equation of motion for the potential $\Phi=x^{2n}/(2n)$, and what's the corresponding coupled first order system?}
\item{Why are there with increasing $n$ phases of constant velocity and a sawtooth pattern in position?}
\item{Please measure the average kinetic and potential energies over many oscillations: What's their ratio $r$ as a function of $n$? Why does $r=\bra T\ket/\bra V\ket$ increase with $n$?}
\end{enumerate}

\begin{figure}[h]
\begin{center}
\resizebox{7.5cm}{!}{\includegraphics{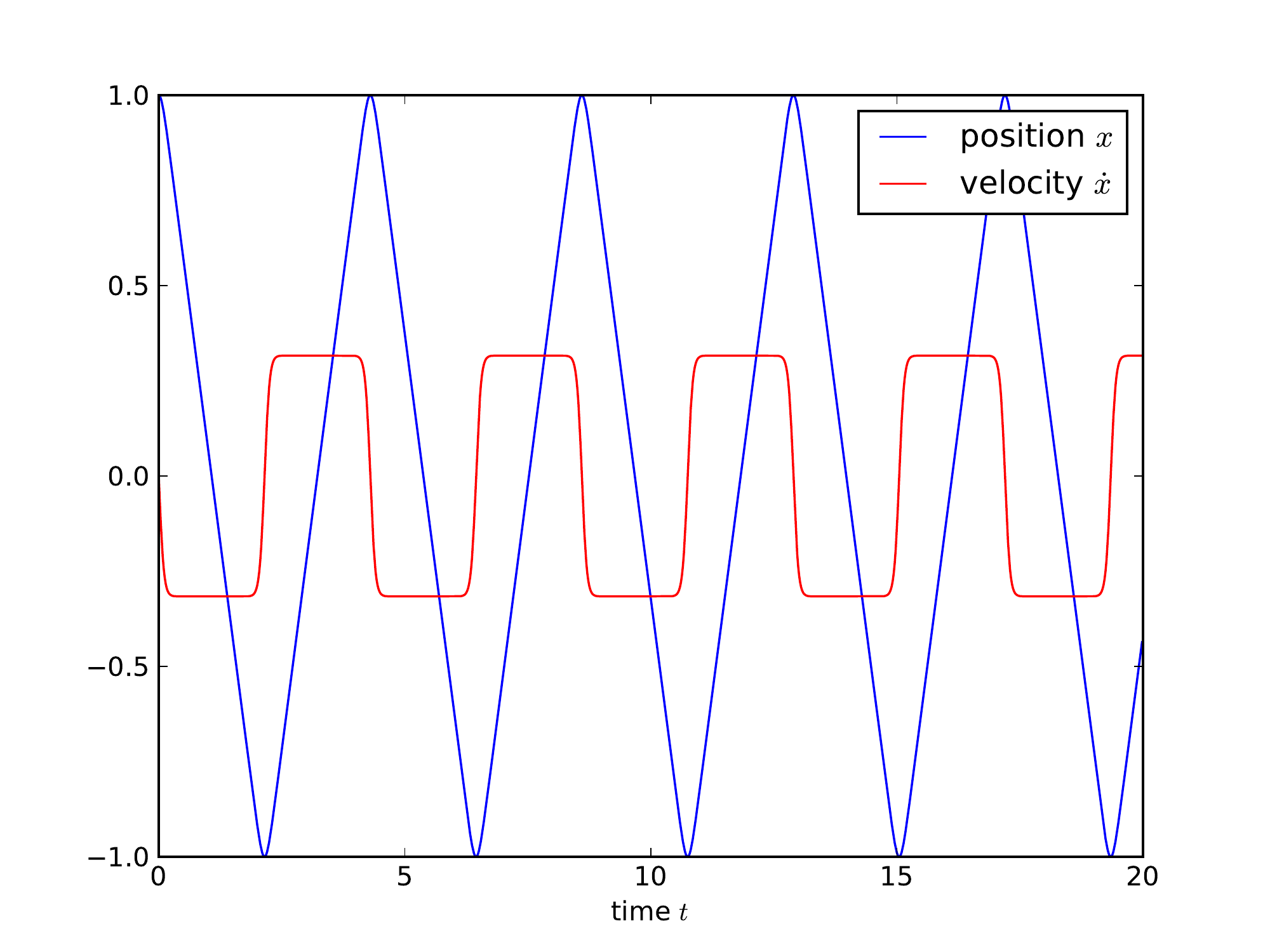}}
\caption{numerical solutions to the anharmonic oscillator for $n=10$}
\end{center}
\end{figure}

\item \question{Virial theorem in the (generalised) Kepler-problem}\\
Please simulate Kepler orbits with the script \path{kepler.py}: The equation of motion of a particle in the potential $\Phi\propto 1/r^\alpha$ can be derived from the Lagrange density
\begin{equation}
\mathcal{L} = \frac{1}{2}\left(\dot{r}^2+r^2\dot{\varphi}^2\right) + \frac{GM}{r^\alpha}
\end{equation}
Please derive the equation of motion with the Euler-Lagrange-equations and convert the resulting second-order equations in a set of first coupled order equations.

\begin{enumerate}[(a)]
\item{Change the total energy of the planet by setting $\delta$ to a value unequal to 0 and observe the change in the orbit. The product constants $GM$ is set to $GM=1$ for the numerics.}
\item{Change the value of $\alpha$ to a number different from 1: Are the orbits still closed? NB: The problem becomes unstable if $\alpha$ is too large, try to experiment in the range $\alpha=0.8\ldots1.2$.}
\item{Can you generate precession motion and orbits lagging behind by choosing a suitable $\alpha$?}
\item{Is it possible to have bound systems for $\alpha\geq 2$? What's the physical reason?}
\item{What's the virial relationship and how does it depend on $\alpha$?}
\end{enumerate}

\begin{figure}[h]
\begin{center}
\resizebox{7.5cm}{!}{\includegraphics{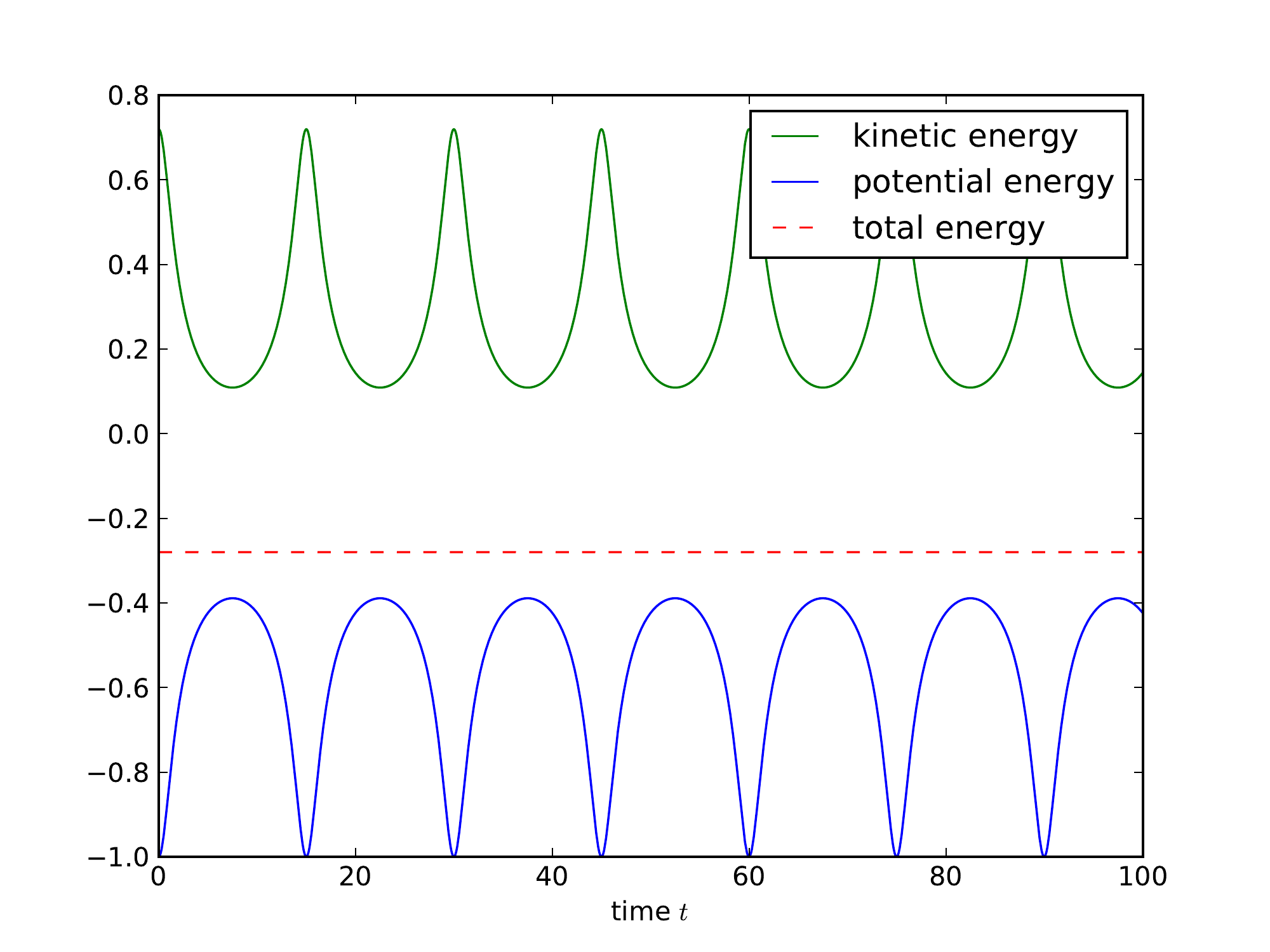}}
\caption{kinetic and potential energy in Kepler-orbits}
\end{center}
\end{figure}

\end{enumerate}

\newdocument{Questions}





\begin{center}
\LARGE \textbf{Astronomy from 4 Perspectives: the Dark Universe}
\HRule
\end{center}
\begin{flushright}
prepared by: Heidelberg participants
\end{flushright}
\begin{center}
{\Large \textbf{Questions: Dark matter and the virial theorem}}
\end{center}
\vspace{5mm}

\begin{enumerate}[\itshape \bfseries 1.]

\item \question{Theory behind the virial theorem}
\begin{enumerate}[(a)]
\item{What's the idea of the virial theorem?}
\item{What's the difference to energy conservation?}
\item{For what kind of system can you use the virial theorem?}
\end{enumerate}

\item \question{Mechanical similarity}
\begin{enumerate}[(a)]
\item{Why is mechanical similarity restricted to potentials of the shape $\Phi\propto r^\alpha$?}
\item{Does $\alpha$ have to be integer?}
\item{What's the generalisation of the Kepler-law for a potential of the form $\Phi\propto r^\alpha$?}
\end{enumerate}

\item \question{Harmonic oscillator}
\begin{enumerate}[(a)]
\item{What can you say about different energy types in the harmonic oscillator on average?}
\end{enumerate}

\item \question{Kepler problem and planetary motion}
\begin{enumerate}[(a)]
\item{Can you show that Kepler's third law implies that potential and kinetic energy are proportional to each other?}
\item{Is this true in general?}
\end{enumerate}

\item \question{Virial theorem in galaxies}\\
\begin{enumerate}[(a)]
\item{Could one explain the rotation curves by assuming a different gravitational law?}
\end{enumerate}

\end{enumerate}

\addcontentsline{toc}{section}{Rotation Curves}
\newdocument{Exercises}





\begin{center}
\LARGE \textbf{Astronomy from 4 Perspectives: the Dark Universe}
\HRule
\end{center}
\begin{flushright}
prepared by: Jena participants
\end{flushright}
\begin{center}
{\Large \textbf{Exercise: Dark matter and galaxy rotation curves}}
\end{center}
\vspace{5mm}

\begin{enumerate}[\itshape \bfseries 1.]

\item \question{Harmonic oscillator and energy types}\\
The harmonic oscillator is described the the differential equation $\ddot{x} = -g/l\: x$, and performs harmonic oscillations $x(t)\propto\exp(\pm\ci\omega t)$ with $\omega^2 = g/l$.
\begin{enumerate}[(a)]
\item{Please show that $\bra T\ket = \bra V\ket$ with the kinetic energy $T$ and the potential energy $V$. The brackets $\bra\ldots\ket$ are time averages over one oscillation period $\tau$,
\begin{equation}
\bra T\ket = \frac{1}{\tau}\int_0^\tau\dd t\:T(t)
\quad\mathrm{and}\quad
\bra V\ket = \frac{1}{\tau}\int_0^\tau\dd t\:V(t)
\end{equation}
which is defined as $\tau = 2\pi/\omega$, and the specific energies $T(t) = \dot{x}^2/2$ and $V(t) = gx^2$/(2l).}
\item{Could you predict the proportionality between $\bra T\ket$ and $\bra V\ket$ from the isochrony of the harmonic oscillator?}
\end{enumerate}
The probability of finding the oscillator at a certain amplitude $x$ is inversely proportional to the velocity: $\dd x/\dd t = \upsilon$, such that $\Delta t = \Delta x/\upsilon$. If the range of motion is divided into equidistant intervals $\Delta x$, the probability $p$ of seeing the oscillator in one of those is proportional to the time it spends there, i.e. proportional to $1/\left|\upsilon\right|$.
\begin{enumerate}[(a)]
\setcounter{enumii}{2}
\item{Please normalise $p$ and draw the function $p(\upsilon)$: If you look randomly at a harmonic oscillator, at what stage in its oscillation are you most likely to see it?}
\item{Please define averages
\begin{equation}
\bra T\ket = \int\dd\upsilon\:p(\upsilon)T(\upsilon)
\quad\mathrm{and}\quad
\bra V\ket = \int\dd\upsilon\:p(\upsilon)V(\upsilon)
\end{equation}
and compute both integrals. You can use energy conservation for the second integral to express $V$ in terms of the velocity $\upsilon$. Are the results identical to the previous computation? Be careful to take the positive sign of $p$ into account, by using the symmetry of the integrand.
}
\item{Why is there no issue with convergence when the probability density $p\rightarrow\infty$ at $\upsilon\rightarrow0$?}
\item{Is the virial relation $\bra T\ket = \bra V\ket$ as well valid for a circular orbit in a spherically symmetric harmonic potential?}
\item{Is it valid as well for any other Lissajous-figure?}
\end{enumerate}

\item \question{Flat rotation curves}\\
Let's consider the motion of stars inside a galaxy with the density profile of a {\em singular isothermal sphere}, which is $\rho\propto r^{-2}$. The singular isothermal sphere describes the density of dark matter well on scales of the galactic disc.
\begin{enumerate}[(a)]
\item{Please show by solving the Poisson equation $\Delta\Phi = 4\pi G\rho$,
\begin{equation}
\Delta\Phi = \frac{1}{r^2}\frac{\dd}{\dd r}\left(r^2\frac{\dd\Phi}{\dd r}\right) = 4\pi G\rho,
\end{equation}
for a spherically symmetric density profile $\rho\propto r^{-2}$ that rotation curves are flat.}
\item{Please compute the mean kinetic $\bra T\ket$ and mean potential energy $\bra V\ket$ for the circular motion in an isothermal sphere as a function of $r$.}
\item{Is it possible in this case to decompose the circular orbiting motion into two uncoupled orthogonal harmonic oscillations?}
\item{What would the density profile need to be such that stars would perform harmonic oscillations through the centre of the galaxy, i.e. for the potential to be quadratic, $\Phi\propto r^{2}$?}
\end{enumerate}

\item \question{MoND, the Solar system and the Milky Way}\\
Let's assume that we can change the acceleration due to gradients in the gravitational potential $\nabla\Phi$ in an empirical way,
\begin{equation}
\frac{\dd\Phi}{\dd r} \rightarrow \frac{\dd\Phi}{\dd r} + a_0,
\end{equation}
as it would be relevant for a circular motion around the Milky Way centre in a spherically symmetric potential.
\begin{enumerate}[(a)]
\item{What would be the effect on a rotation curve from the density profile $\rho\propto r^{-\alpha}$?}
\item{The parameter $a_0$ would need to be chosen small: Please estimate an upper bound on the value of $a_0$ from the orbital acceleration of the Solar system on its passage around the Milky Way center. You can find all necessary data on Wikipedia.}
\item{Please think of a way to visualise the numerical value of $a_0$.}
\item{At what distance from the Earth's surface would the gravitational acceleration be $a_0$?}
\end{enumerate}

\end{enumerate}

\newdocument{Play with data}






\begin{center}
\LARGE \textbf{Astronomy from 4 Perspectives: the Dark Universe}
\HRule
\end{center}
\begin{flushright}
prepared by: Jena participants and Bj{\"o}rn Malte Sch{\"a}fer
\end{flushright}
\begin{center}
{\Large \textbf{Play with data: Rotation curves of galaxies}}
\end{center}
\vspace{5mm}

\noindent
Observations of the rotation of disc galaxies is nowadays done in the H$\alpha$-line of hydrogen, because it reaches to much larger distances from the galaxy centre compare to the stellar light. In this exercise we have a look at a data set on low surface-brightness galaxies by W. de Blok, S. McGaugh and V. Rubin, Astronomical Journal 122, 2381 (2001).

\begin{enumerate}[\itshape \bfseries 1.]

\item \question{Flat rotation curves}\\
Let's start by exploring H$\alpha$-data for low surface-brightness galaxies.
\begin{enumerate}[(a)]
\item{Why is it a clever idea to focus on galaxies with a low (optical) surface brightness?}
\item{What is the general relationship between rotation curve $\upsilon(r)$ and mass profile $\rho(r)$?}
\item{A isothermal sphere with a core has the density profile $\rho(r)$,
\begin{equation}
\rho(r) = \rho_0\left(1+\left(\frac{r}{r_c}\right)^2\right)^{-1},
\end{equation}
with the central density $\rho$ and the core radius $r_c$. The corresponding velocity profile $\upsilon(r)$,
\begin{equation}
\upsilon(r)^2 = 4\pi G\rho_0 r_c^2\left(1-\frac{r_c}{r}\arctan\left(\frac{r}{r_c}\right)\right),
\end{equation}
with the gravitational constant $G\simeq 10^{-11}~\mathrm{m}^3/\mathrm{kg}/\mathrm{s}^2$.}
\item{Please show that the asymptotic value for $\upsilon$ for $r\rightarrow\infty$ is $\upsilon_\infty = \sqrt{4\pi G\rho_0r_c^2}$. Please check the units of the relation between $\upsilon_\infty$ and $r_c$ and $\rho_0$.}
\item{Please use the script \path{rotplot.py} and plot a couple of rotation curves: Do they show the expected behaviour?}
\end{enumerate}

\begin{figure}[h]
\begin{center}
\resizebox{9cm}{!}{\includegraphics{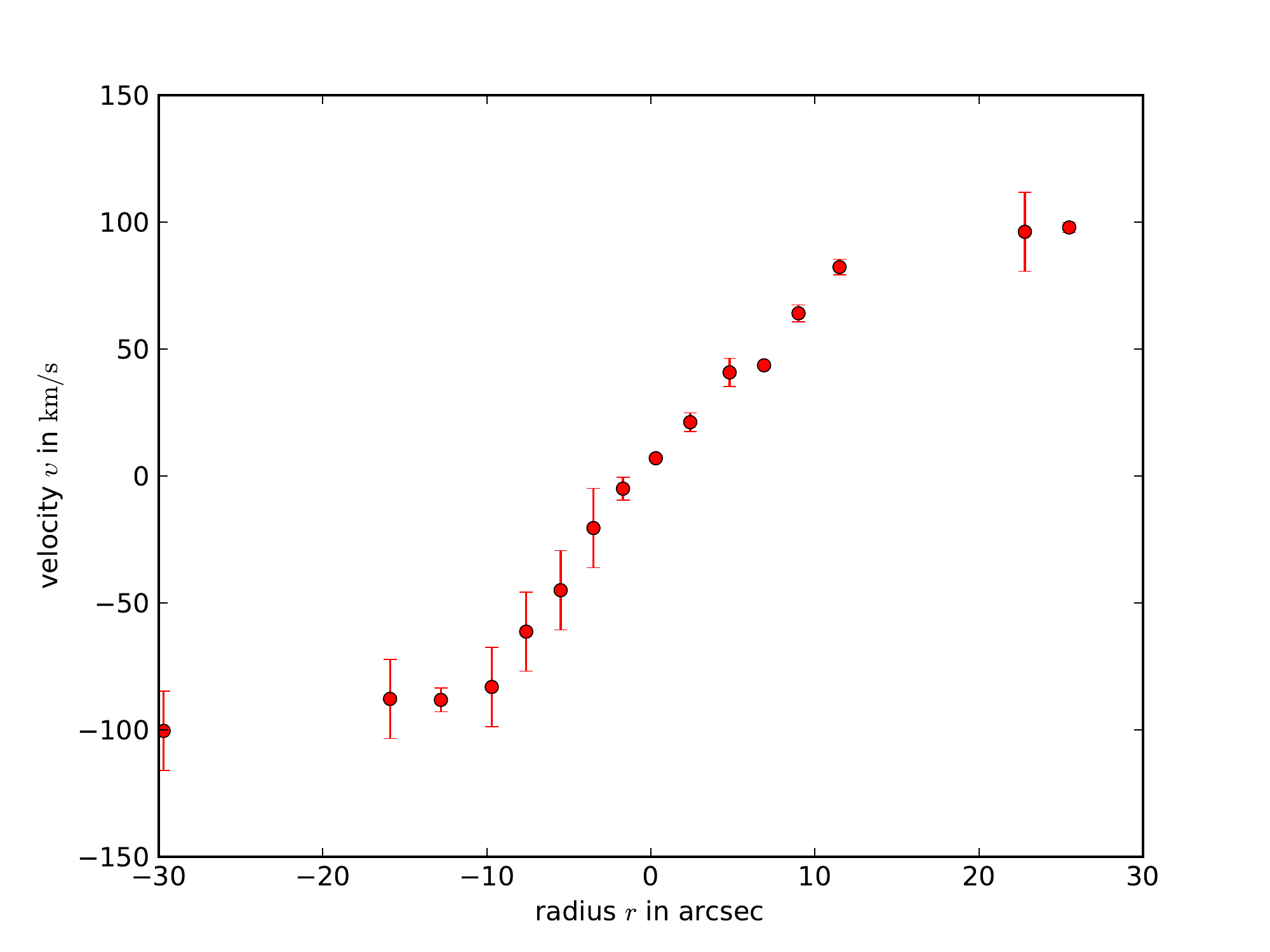}}
\caption{rotation curve $\upsilon(r)$ of the galaxy F568}
\end{center}
\end{figure}

\item \question{Luminous and dark matter}\\
With the script \path{rotfit.py} you can fit a model rotation curve to data. Take care to read off the distance $d$ to the galaxy from the table, in order to convert $r$ from arc seconds to $\mathrm{kpc}$.
\begin{enumerate}[(a)]
\item{Are the curves from the isothermal-sphere model providing a good fit to data?}
\item{What are typical velocities $\upsilon_\infty$, central densities $\rho_0$ and core radii $r_c$?}
\item{What is the role of $\epsilon$ in the script? Why are the results not affected if $\epsilon$ is small enough?}
\end{enumerate}
Please continue by completing the table.
\begin{enumerate}[(a)]
\setcounter{enumii}{3}
\item{Please try to find out if the mass to light-ratio $M/L$ is large: For that purpose, estimate the total mass $M$ in units of the solar mass $M_\odot = 10^{30}~\mathrm{kg}$,
\begin{equation}
M = 4\pi\int_0^\infty r^2\mathrm{d}r\: \rho(r),
\end{equation}
and compare it to the total luminosity. For the integral, you can use the result
\begin{equation}
\int\dd x\:\frac{x^2}{1+x^2} = \arctan(x) + \mathrm{const}.
\end{equation}
Please truncate the integration at the tidal radius $10r_c$. With the expression for the mass, please verify the relationship between orbital velocity $\upsilon$ and distance $r$.
}
\item{Then, please express the mass to light-ratio $M/L$ in units of solar masses per solar luminosities $M_\odot/L_\odot$: The luminosity $L$ in units of the solar luminosity $L_\odot$ follows from the difference of the absolute magnitudes,
\begin{equation}
\frac{L}{L_\odot} = 10^{0.4(\mathrm{Mag}_\odot - \mathrm{Mag})},
\end{equation}
you can find the values for $\mathrm{Mag}$ of the galaxies in the table, and use the literature value for $\mathrm{Mag}_\odot=5.45$ in the same band ($R$-band) from the literature.
}
\item{Is there evidence for dark matter?}
\end{enumerate}

\begin{figure}[h]
\begin{center}
\resizebox{9cm}{!}{\includegraphics{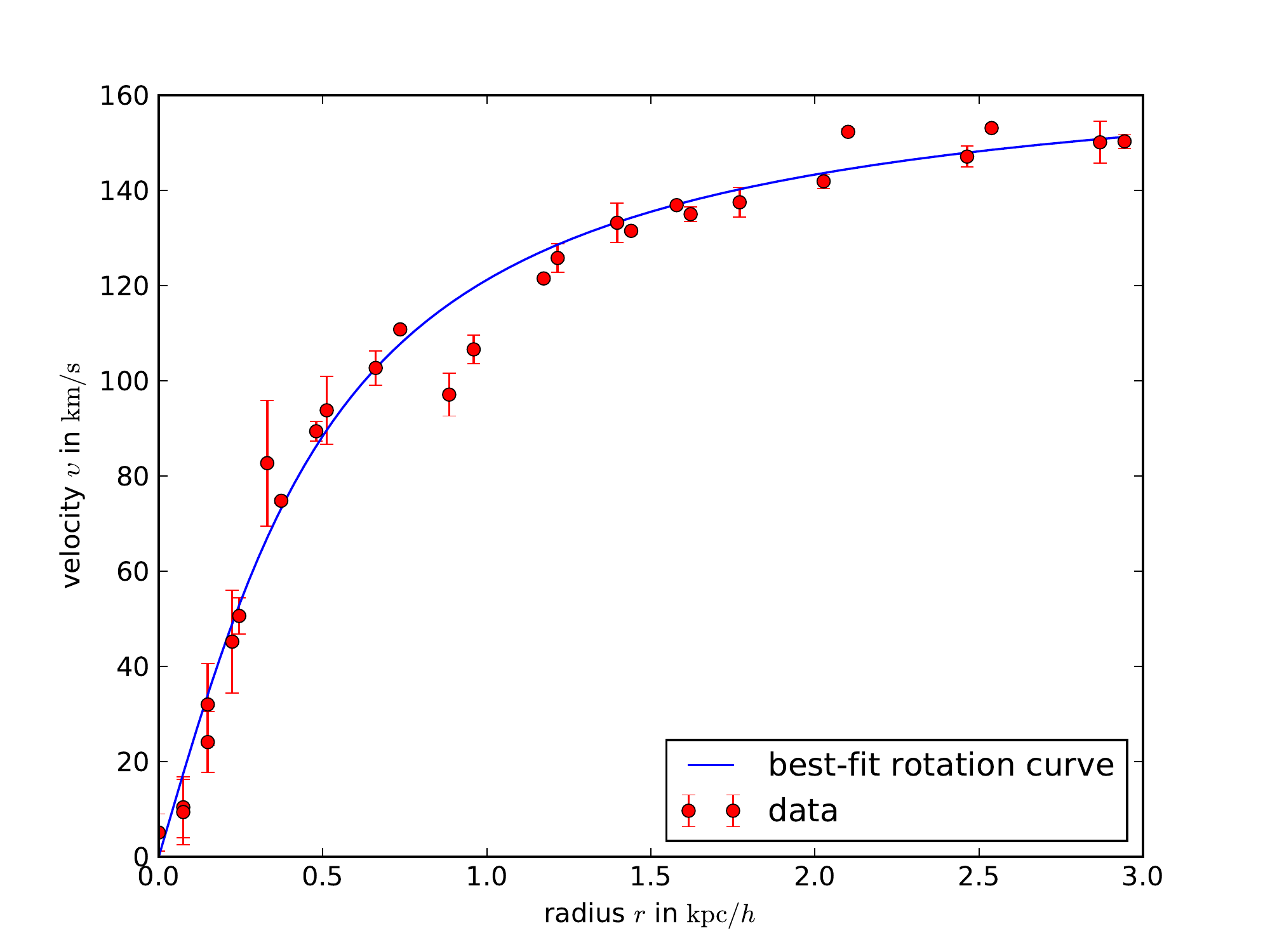}}
\caption{fit of a isothermal sphere rotation curve $\upsilon(r)$ to the galaxy U11557 at 22 Mpc distance, with the values $r_c=0.412~\mathrm{kpc}$ and $\upsilon_\infty = 169~\mathrm{km}/\mathrm{s}$.}
\end{center}
\end{figure}

\begin{table}[h]
\begin{center}
\begin{tabular}{|l|r|r|r|r|r|r|r|}
\hline
galaxy		& $d$ in $\mathrm{Mpc}$	& $\upsilon_\infty$ in $\mathrm{km}/\mathrm{s}$ & $r_c$ in $\mathrm{kpc}$ & $M$ in $M_\odot$ & $\mathrm{Mag}$ & $M/L$ in $M_\odot/L_\odot$\\
\hline
E0140040	& 212	& & & & -21.6 & \\
E0840411	& 80	& & & & -18.1 & \\
E1200211	& 15	& & & & -15.6 & \\
E1870510	& 18	& & & & -16.5 & \\
E2060140	& 60	& & & & -19.2 & \\
E3020120	& 69	& & & & -19.1 & \\
E3050090	& 11	& & & & -17.3 & \\
E4250180	& 86	& & & & -20.5 & \\
E4880049	& 22	& & & & -16.8 & \\
F563-1		& 45	& & & & -17.3 & \\
F568-3		& 77	& & & & -18.3 & \\
F571-8		& 48	& & & & -17.6 & \\
F579-V1		& 85	& & & & -18.8 & \\
F583-1		& 32	& & & & -16.5 & \\
F583-4		& 49	& & & & -16.9 & \\
U4115		& 3.2	& & & & -12.4 & \\
U5750		& 56	& & & & -18.7 & \\
U6614		& 85	& & & & -20.3 & \\
U11454		& 91	& & & & -18.6 & \\
U11557		& 22	& 169 & 0.412 & & -20.0 & \\
U11583		& 5		& & & & -14.0 & \\
U11616		& 73	& & & & -20.3 & \\
U11648		& 48	& & & & -21.0 & \\
U11748		& 73	& & & & -22.9 & \\
U11819		& 60	& & & & -20.3 & \\
\hline
\end{tabular}
\end{center}
\end{table}

\end{enumerate}
\newdocument{Questions}





\begin{center}
\LARGE \textbf{Astronomy from 4 Perspectives: the Dark Universe}
\HRule
\end{center}
\begin{flushright}
prepared by: Jena participants
\end{flushright}
\begin{center}
{\Large \textbf{Questions: Dark matter and galaxy rotation curves}}
\end{center}
\vspace{5mm}

\begin{enumerate}[\itshape \bfseries 1.]
\item \question{Orientations of galaxies}\\
Think about how the galaxy should be orientated to be observed?\\
Here are some pictures as example:
\begin{figure}[h!]
	\centering
	\begin{subfigure}[t]{0.3\textwidth}
	\includegraphics[height=5cm]{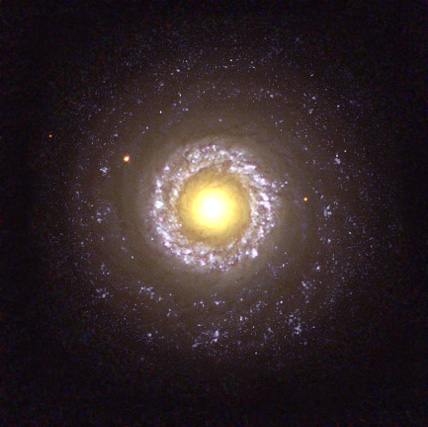}
	\caption{Edge on galaxy NGC 7742\\Inclination angle $i =0^\circ$\\Image credit:\\Hubble Heritage Team\\(AURA/STScI/NASA/ESA)}
	\end{subfigure}
	\begin{subfigure}[t]{0.3\textwidth}
	\includegraphics[height=5cm]{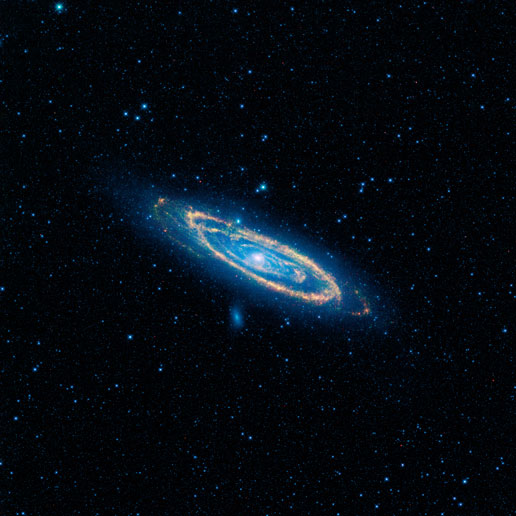}
	\caption{Our galactic neighbour Andromeda as seen in infrared.\\
	Inclination angle $i \approx 13^\circ$\\Image credit:\\NASA/JPL-Caltech/UCLA}
	\end{subfigure}
	\begin{subfigure}[t]{0.3\textwidth}
	\includegraphics[height=5cm]{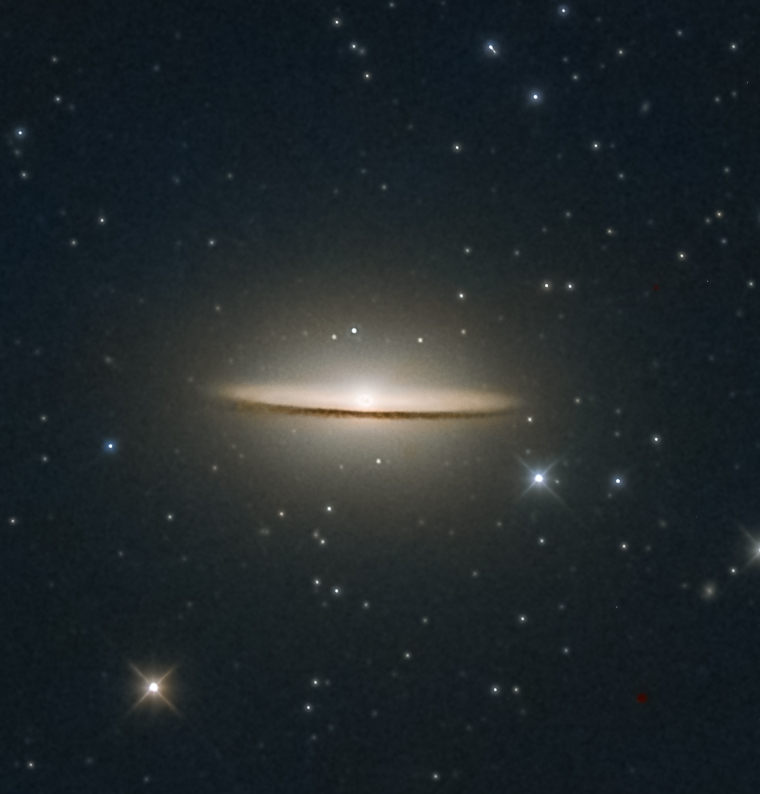}
	\caption{The almost edge-on sombrero galaxy.\\
	Inclination angle $i \approx 90^\circ$\\Image credit:\\Carsten Frenzl}
	\end{subfigure}
\end{figure}

\item \question{Galactic Rotation curves}\\
Calculate the radial velocities from the measured wavelengths and plot them over the distance from the galaxy center. use $\lambda_0 = 21.106\,$\AA\, and 1\,pc = $3.1\cdot 10^{16}$ m.
\begin{table}[h]
\centering
\begin{tabular}{c|c|c}
$\lambda$ in \,\AA &Radius $R$ in Mpc& $v_{\text{rotation}}$ in $\frac{km}{s}$\\\hline
21.1195&1& \\
21.1130&2&\\
21.1173&5&\\
21.1194&7&\\
21.1201&10&\\
21.1208&15&\\
21.1211&20&\\
21.1215&22&\\
21.1213&25&\\
\end{tabular}
\end{table}

\item \question{Circular obits}\\
Derive for circular orbits the formula for the velocity $v$ in dependence of the distance $r$. Assume a radially symmetric mass distribution.
\item \question{Velocity of planets}\\
Assuming circular orbits, compute the velocities of the planets in our solar system. Plot the resulting rotation curve $v$ over $r$.
\item \question{Expected Rotation curve}\\
Formulate an expectation for the rotation curve of the Milky Way, assuming the mass in the bulge to be $1.6\cdot 10^{10}\,\text{M}_\odot$ and the disk to be $4\cdot 10^{10}\,\text{M}_\odot$
\item \question{Observed rotation curve}
The observed rotation curves of spiral galaxies are of the following form:
\begin{figure}[h]
	\centering
	\includegraphics[width=0.5\textwidth]{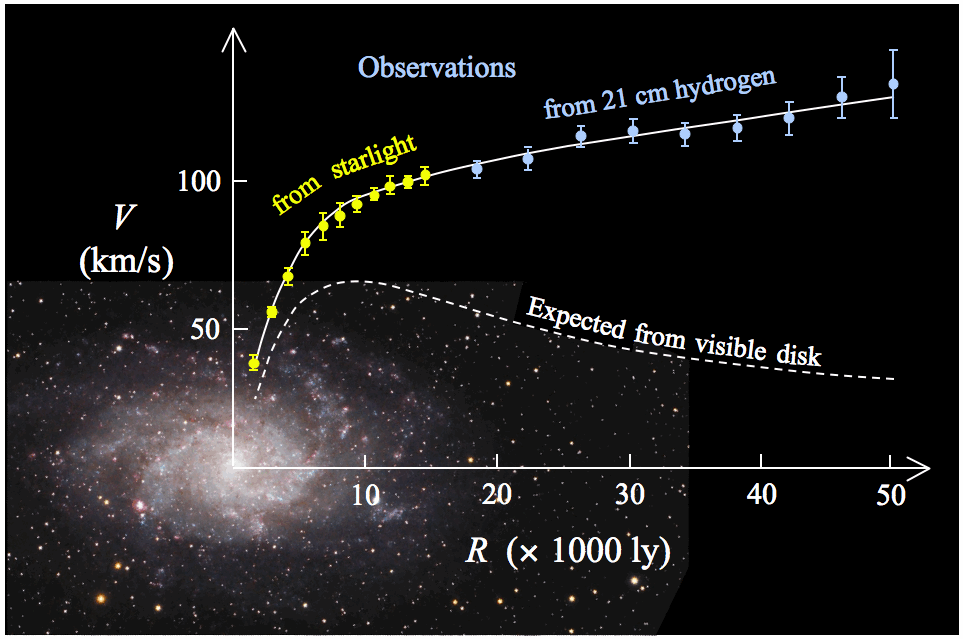}
	\caption{Expected and observed rotation curve of M33.\\Image credit: Steffania Deluca}
\end{figure}
This cannot be explained by visible mass alone. Assuming that dark matter is the source of the difference between the observed and the predicted rotation curves, please calculate the mass of the dark matter depending on the velocities $v(r)$ and $v_{\text{axis}}(r)$
\item \question{Dark matter distribution}
To find out how dark matter is distributed throughout a spiral galaxy, please consider a simple rotation curve consisting of a linear and a constant branch. Assume a spherically symmetric mass distribution of the form 
\begin{equation*}
\mu(r) \propto r^k
\end{equation*}
For the mass use the formula
\begin{equation*}
M(r) = 4\pi\int_{r_0}^{r}\mu(\rho)\,\rho^2 \text{d}\rho
\end{equation*}
	\begin{enumerate}
	\item Please calculate the mass of the bulge in dependence of $k$
	\item Using the formula for $v$ from Task 3 and the result from a), please determine the exponent $k$ for the bulge. Calculate the mass of the bulge in dependence of $r$ and the complete mass $M_B$ of the bulge.
	\item To determine $k$ for the halo ($v=$ const.), consider the total mass to be composed of the mass of the bulge $M_B$ and the mass of the halo $M_H$.
	\begin{equation*}
		M(r) = M_B +M_H(r)
	\end{equation*}
	Calculate the mass outside of the bulge with the integral for the mass. Determine the exponent $k$ using the results of b) and Task 3. Find a formula for the mass of the halo in dependence of $r$.
	\item Compare the rotation curve of the bulge to the rotation curve of a rigid body.
	\end{enumerate}
\end{enumerate}

\addcontentsline{toc}{section}{Planck and the Cosmic Microwave Background (CMB)}
\newdocument{Exercises}





\begin{center}
\LARGE \textbf{Astronomy from 4 Perspectives: the Dark Universe}
\HRule
\end{center}
\begin{flushright}
prepared by: Padova participants and Bj{\"o}rn Malte Sch{\"a}fer
\end{flushright}
\begin{center}
{\Large \textbf{Exercise: Planck-spectrum and the CMB}}
\end{center}
\vspace{5mm}

\begin{enumerate}[\itshape \bfseries 1.]

\item \question{Properties of the Planck-spectrum}\\
Let's derive the fundamental properties of the Planck-spectrum,
\begin{equation}
S(\nu) = S_0\frac{\nu^3}{\exp(h\nu/(k_BT))-1}
\quad\rightarrow\quad
S(\nu) = S_0\nu^3\exp(-h\nu/(k_BT)),
\end{equation}
by using Wien's approximation (the second expression), which makes the integrals easier. The constant $S_0$ depends only on numbers, natural and mathematical constants.
\begin{enumerate}[(a)]
\item{Please compute the total intensity $\int_0^\infty\dd\nu\:S(\nu)$ and show that it is $\propto T^4$.}
\item{Show that the position $\nu_m$ of the maximum scales $\nu_m\propto T$.}
\item{Please derive the scaling of the mean
\begin{equation}
\bra\nu\ket = \frac{\int_0^\infty\dd\nu\:\nu S(\nu)}{\int_0^\infty\dd\nu\:S(\nu)}
\end{equation}
and show that it is proportional to $T$.}
\item{Is there a fixed ratio between $\bra\nu\ket$ and $\nu_m$?}
\item{In which limit is Wien's approximation applicable?}
\item{Do the scaling behaviours with $T$ derived above depend on the details of the distribution?}
\end{enumerate}

\item \question{Wien's distribution function}\\
Let's stick for a second with Wien's distribution function in $n$ dimensions,
\begin{equation}
S(\nu) = S_0\nu^n\exp(-h\nu/(k_BT)),
\end{equation}
and derive a few general properties, which will hold for the Planck-distribution as well (although the computations are more complicated).
\begin{enumerate}[(a)]
\item{Please begin by showing that
\begin{equation}
\int_0^\infty\dd x\:x^n\exp(-x) = n!
\end{equation}
using $n$-fold integration by parts.}
\item{Alternatively, please show the recursion relation of the $\Gamma$-function, 
\begin{equation}
\Gamma(n) = (n-1)\Gamma(n-1),
\quad\mathrm{together~with}\quad
\Gamma(0) = 1.
\end{equation}
The $\Gamma$-function is defined by 
\begin{equation}
\Gamma(n) = \int_0^\infty\dd x\:x^{n-1}\exp(-x),
\end{equation}
and is related to the factorial by $\Gamma(n) = (n-1)!$}
\item{What scaling of the moments
\begin{equation}
\bra\nu^m\ket = \frac{\int_0^\infty\dd\nu\:\nu^mS(\nu)}{\int_0^\infty\dd\nu\:S(\nu)}
\end{equation}
with temperature $T$ do you expect?
}
\item{Please show that the skewness parameter $s = \bra\nu^3\ket/\bra\nu^2\ket^{3/2}$ and the kurtosis parameter $k = \bra\nu^4\ket/\bra\nu^2\ket^2$ are independent from the temperature $T$. What would be the physical interpretation of $s$ and $k$?}
\item{Would an equivalent result be true for the parameter $\bra\nu^{2n}\ket/\bra\nu^2\ket^n$?}
\end{enumerate}

\item \question{Planck-spectra at cosmological distances}\\
Imagine you observe an object emitting a Planck-spectrum at a cosmological distance, such that all photons arrive with a redshifted frequency $\nu\rightarrow a\nu=\nu / (1+z)$ with scale factor $a$ (remember $a<1$) and redshift $z$. A couple of students discusses the fact that the temperature scales with $T\propto 1/a$ and that the photons are redshifted: What's your opinion on the different arguments?
\begin{enumerate}[(a)]
\item{Johannes from Heidelberg says: The temperature $T$ of a photon gas is linked to the thermal energy $E$ by $E=k_BT$. Then, the relativistic dispersion relation of the photons assumes $E = cp$ with the momentum $p$. The momentum $p$ is given by the de Broglie-relationship as $p=h/\lambda$. If now the photon wavelength is changed $\lambda\rightarrow a\lambda$, the temperature needs to scale $T\propto 1/a$.}
\item{Antonia from Padova says: What about a purely thermodynamical argument? A gas of photons has an adiabatic index of $\kappa=4/3$, and the Hubble expansion is an adiabatic change of state, because there is no thermal energy created or dissipated. Then, the adiabatic invariant says that $TV^{\kappa-1}$ is conserved, which gives me $T\propto 1/a$ with $V\propto a^3$. And I understand why entropy is conserved but not energy.}
\item{Marlene from Jena says: Due to the Hubble-expansion, every point is in recession motion with respect to every other point. If a photon gets scattered into your direction, the scattering particle will necessarily move away from you, leading to a lower perceived energy and a larger wavelength. It's important to view it like that because a photon gas can not change its state without interaction due to the linearity of electrodynamics, and this argument shows that it's a kinematic effect: It's a similarity transform of the Planck-spectrum.}
\item{Lorenzo from Florence says: It's important for the Planck-spectrum that the mean particle separation and the thermal wavelength are identical. You can only arrange for that if $T\propto 1/a$ if the particle separation increases $\propto a$. I'm only assuming that the number of photons is conserved, but not their energy, and that everything stays in equilibrium.}
\end{enumerate}

\item \question{CMB as a source of energy}\\
A Carnot-engine converts thermal energy taken from two reservoirs at different temperatures into mechanical energy at the efficiency $\eta = 1-T_2/T_1$.
\begin{enumerate}[(a)]
\item{Estimate if one can use the temperature anisotropies in the CMB of $\Delta T/T\simeq 10^{-5}$ to generate mechanical energy. How much power could you realistically generate? Construct a machine that converts radiation power into mechanical energy.}
\item{Could you use the time evolution of the CMB-temperature for this purpose? You know already that $T\propto 1/a$, so please construct a machine that produces energy from the CMB.}
\item{Why does a solar cell transform the radiation from the Sun into electrical energy? One might argue that the Planck-spectrum is that of thermal equilibrium in which case the mechanical work is zero: Due to the first law, mechanical work can not be performed in thermal equilibrium. (please be careful: trick question)}
\end{enumerate}

\end{enumerate}

\newdocument{Solutions}





	
	\begin{center}
		\LARGE \textbf{Astronomy from 4 Perspectives: the Dark Universe}
		\HRule
	\end{center}
	\begin{flushright}
		prepared by: Padua participants
	\end{flushright}
	\begin{center}
		{\Large \textbf{Exercise: Planck spectrum and CMB}}\\
		\vspace*{2mm}
		{\Large \textbf{Solutions}}
		
	\end{center}
	\vspace{5mm}
	
	\begin{enumerate}[\itshape \bfseries 1.]
	\item \question{Properties of the Planck Spectrum}\\
	\begin{align}
	S(\nu) =S_0\frac{\nu^3}{\enkt -1} \Rightarrow S(\nu)=S_0\nu^3 \enkt
	\end{align}
	
	\begin{enumerate}[(a)]
	\item
		\begin{align}
			\intzinf S(\nu)=\intzinf S_0 \nu^3 \enkt
		\end{align}
		\begin{align}
			x=\frac{h\nu}{kT}\quad;\quad \nu=\frac{kT}{h}x \quad ;\quad d\nu=\frac{kT}{h}dx
		\end{align}
		\begin{align}
			\intzinf S_0\left(\frac{kT}{h}\right)^3\enkt\left(\frac{kT}{h}\right) \propto T^4
		\end{align}
	\item
	\begin{align}	
	 \frac{dS}{d\nu}=0\Rightarrow \frac{dS}{d\nu}=&3S_0\nu^2 \enkt+S_0\nu^3\left(-\frac{h}{kT}\right)\enkt = 
	 \\&=S_0\nu^3\enkt\left(3-\frac{h\nu}{kT}\right)
	\end{align}
	\begin{align}
	3-\frac{h\nu}{kT}=0 \Rightarrow \nu=\frac{3kT}{h}\propto T
	\end{align}
	\item
	\begin{align}
	<\nu>=\frac{\intzinf\nu S(\nu)d\nu}{\intzinf S(\nu)d\nu}= \frac{\intzinf \nu^3 S(\nu)d\nu}{\intzinf S(\nu)d\nu}
	\end{align}
	\begin{align}
	x=\frac{h\nu}{kT} \quad ;\quad \nu = \frac{kT}{h}x \quad d\nu = \frac{kT}{h}dx
	\end{align}
	\begin{align}
	\frac{\intzinf S_0 \frac{kT^4}{h}\emx\frac{kT}{h}dx}{\intzinf S_0 \frac{kT^3}{h}\emx\frac{kT}{h}dx}
	\end{align}
	\begin{align}
	S\propto \frac{T^5}{T^4}\propto T
	\end{align}
	\item Yes, as both of them depend on T
	\item Only at high energies
	\item No, as if we consider any power of T in the exponential, any substitution $x=\frac{h\nu}{kT}$ gives a new differential $d\nu\frac{kT^4}{h}$, both in numerator and denominator
	\end{enumerate}
	\item \question{Wien's distribution function}
		\begin{align}
		S(\nu) = S_0\nu^n \enkt \quad;\quad\Gamma(n) = \intzinf dx\,x^{n-1}\emx
		\end{align}
		\begin{enumerate}[(a)]
		\item 
			\begin{align}
			\intzinf dx\,x^n \e^{-x} = \cancel{-[e^{-x}x^n]^\infty_0} -n\intzinf dx\,x^{n-1}e^{-x}=\intzinf dx\,nx^{n-1}\emx
			\end{align}
			\begin{align}
			=\cancel{x[\emx nx^{n-1}]_0^\infty}-\intzinf dx\,-n(n-1)x^{n-2}=\intzinf dx\, -n(n-1)x^{n-2}\emx
			\end{align}
			\begin{align}
			=[\dots]=
			\end{align}
			\begin{align}
			=n!\intzinf dx\,\emx=n!
			\end{align}
		Let's show that $\Gamma(1)=1$:
			\begin{align}
			\Gamma (1) =\intzinf x^0\emx
			\end{align}
		To demontstrate our relation recursively, let's start by showing it works for n=2:
		\begin{align}
		\Gamma(2) =\intzinf x\emx dx = [-\emx x]_0^\infty - \intzinf dx\, (-\emx) = 1
		\end{align}
		\begin{align}
		=(2-1)\Gamma(1)=1
		\end{align}
		We can now show it for $n+1$:
		\begin{align}
		\Gamma(n+1)=n\,\Gamma(n) = n\,(n-1)\,\Gamma(n-1)=n(n-1)\intzinf x^{n-2}\emx dx
		\end{align}
		Let's introduce $m=n-2\quad \Rightarrow \quad n=m+2$
		\begin{align}
		\Gamma(n+1)=(m+2)(m+1)\intzinf x^m\emx dx = (m+2)(m+1)m! = (m+2)! = n!
		\end{align}
		But
		\begin{align}
			\Gamma(n+1)=\intzinf x^n \emx dx = n!
		\end{align}
	\item
		\begin{align}
		< \nu^m>=\frac{\intzinf \nu^mS(\nu)d\nu}{\intzinf S(\nu)d\nu}
		\end{align}
		\begin{align}
		=\frac{\intzinf\nu^mS_0\nu^n \enkt d\nu}{\intzinf S_0 \nu^n\enkt d\nu}
		\end{align}
		\begin{align}
		x=\frac{h\nu}{kT}\quad;\quad \nu=\frac{kT}{h}x\quad ;\quad d\nu=\frac{kT}{h}dx
		\end{align}
		\begin{align}
		<\nu^m>=\frac{\intzinf\left(\frac{kT}{h}\right)^mS(\nu) \left(\frac{kT}{h}\right)^n\emx \frac{kT}{h}d\nu}{\intzinf S_0\left(\frac{kT}{h}\right)^n\emx \frac{kT}{h}d\nu}
		\end{align}
		\begin{align}
		\propto \frac{T^{m+n+1}}{T^{n+1}} = T^m
		\end{align}
		$\Rightarrow$ The momenta scale like the power chosen.
	\item
		From the result on point 1(c), we  find:
		\begin{align}
		S=\frac{< \nu^3>}{(<\nu>^2)^{3/2}}\propto \frac{T^3}{(T^2)^{3/2}}
		\end{align}
		\begin{align}
		k=\frac{<\nu^4>^2}{<\nu^2>^2}\propto \frac{T^4}{(T^2)^2}
		\end{align}
		Neither of which depends on T.
		\item Yes, again from the result of point 1(c):
		\begin{align}
		\frac{<\nu^{2n}>}{< \nu^2>^n}\propto \frac{T^{2n}}{(T^2)^n}
		\end{align}
		\end{enumerate}
	\item \question{Planck Spectra at cosmological distances}\\
	The solution to this question was discussed live in the classroom
	\item \question{CMB as a source of energy}
		\begin{enumerate}[(a)]
			\item The answer (a) with the Stephan Boltzmann law, the power per $m^2$ can be produced if:
			\begin{align}
			w=10^{-5}\sigma T^4=3\cdot 10^{-11} \,\textrm{Wm}^{-2}
			\end{align}
		\end{enumerate}
	\end{enumerate}
	
\newdocument{Play with data}






\begin{center}
\LARGE \textbf{Astronomy from 4 Perspectives: the Dark Universe}
\HRule
\end{center}
\begin{flushright}
prepared by: Padova participants and Bj{\"o}rn Malte Sch{\"a}fer
\end{flushright}
\begin{center}
{\Large \textbf{Play with data: Planck spectrum and the CMB}}
\end{center}
\vspace{5mm}

\noindent
The satellite COBE observed the cosmic microwave background from 1989 to 1993. One experiment, FIRAS (Far Infrared Absolute Spectrophotometer), took a very precise measurement of the Planck-shape of the CMB, see D.J. Fixsen et al., Astrophysical Journal 420, 445 (1994).

\begin{enumerate}[\itshape \bfseries 1.]

\item \question{CMB-temperature}\\
In this exercise you can play with COBE-data and explore the properties of the Planck-spectrum.\\
Please have a look at the python-script \path{planck_plot.py}, which reads the data file from COBE and plots flux $S(\nu)$ as a function of frequency $\nu$. In addition, it plots Planck-spectra $S(\nu, T)$ for a given temperature $T$. 
\begin{enumerate}[(a)]
\item{What's your measurement for the CMB-temperature $T_\mathrm{CMB}$?}
\item{In what range can you vary $T$ such that the data is well described by the Planck-curve?}
\end{enumerate}

\begin{figure}[h]
\begin{center}
\resizebox{9cm}{!}{\includegraphics{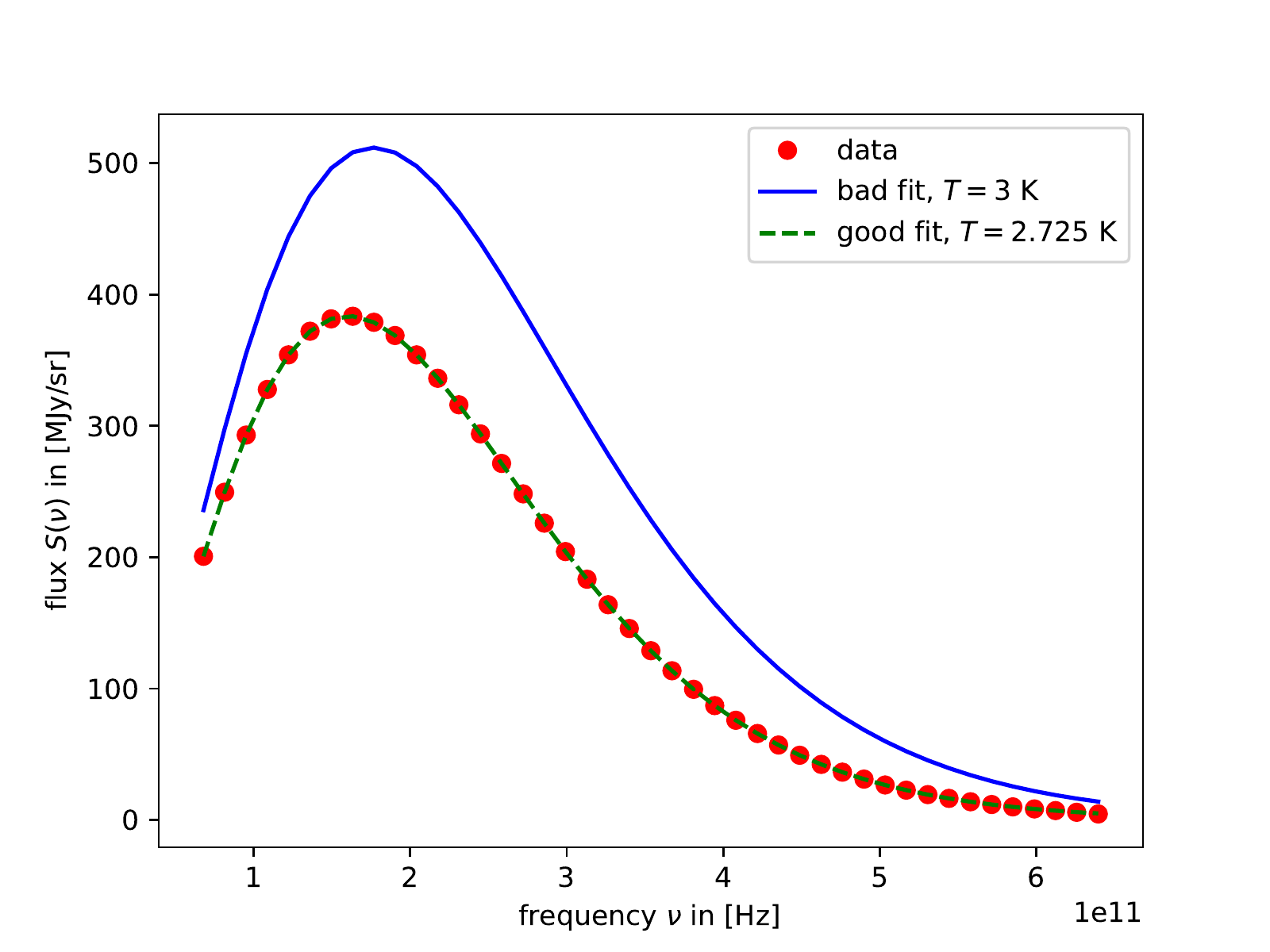}}
\caption{Planck-spectra for different temperatures $T$ superimposed on the COBE-data}
\end{center}
\end{figure}

\item \question{Different radiation laws}\\
The script \path{planck_fit.py} does a proper regression of a model $S(\nu,T)$ to the data, by minimizing the squared difference between data and model, in units of the noise. There are two models for $S(\nu,T)$, the Planck-spectrum and the simplified Wien spectrum.
\begin{enumerate}[(a)]
\item{Carry out a fit to the data with both models: What are the temperatures $T$?}
\item{Which model is better at explaining the data?}
\end{enumerate}

\begin{figure}[h]
\begin{center}
\resizebox{9cm}{!}{\includegraphics{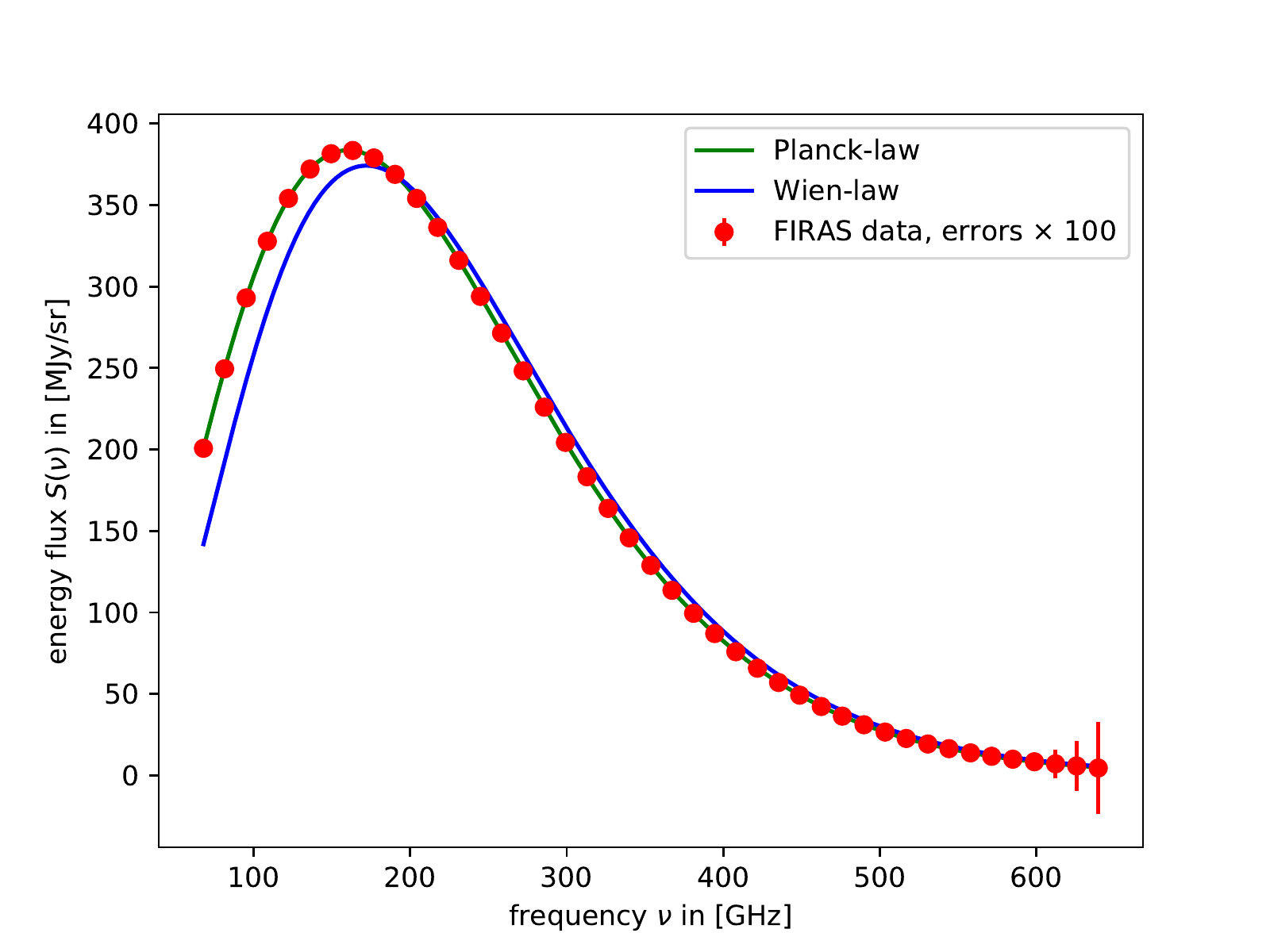}}
\caption{fits of the Planck- and Wien-radiation laws $S(\nu,T)$ to COBE-data}
\end{center}
\end{figure}

\item \question{Precision of the measurement}\\
In running the script \path{planck_likelihood.py} you can estimate which range of values for $T$ would be a good fit. It plots the likelihood $\mathcal{L}(T)\propto \exp(-\chi^2(T)/2)$, with 
\begin{equation}
\chi^2(T) = \sum_{i=1}^{n_\mathrm{data}}\left(\frac{S_i-S(\nu_i,T)}{\sigma_i}\right)^2
\end{equation}
for the $n_\mathrm{data}$ data points $S_i$ at the frequencies $\nu_i$. The statistical error is given by the width of the resulting Gauss-curve. Would it be possible to measure the Planck-constant $\hbar$ parallel to the temperature?

\begin{figure}[h]
\begin{center}
\resizebox{9cm}{!}{\includegraphics{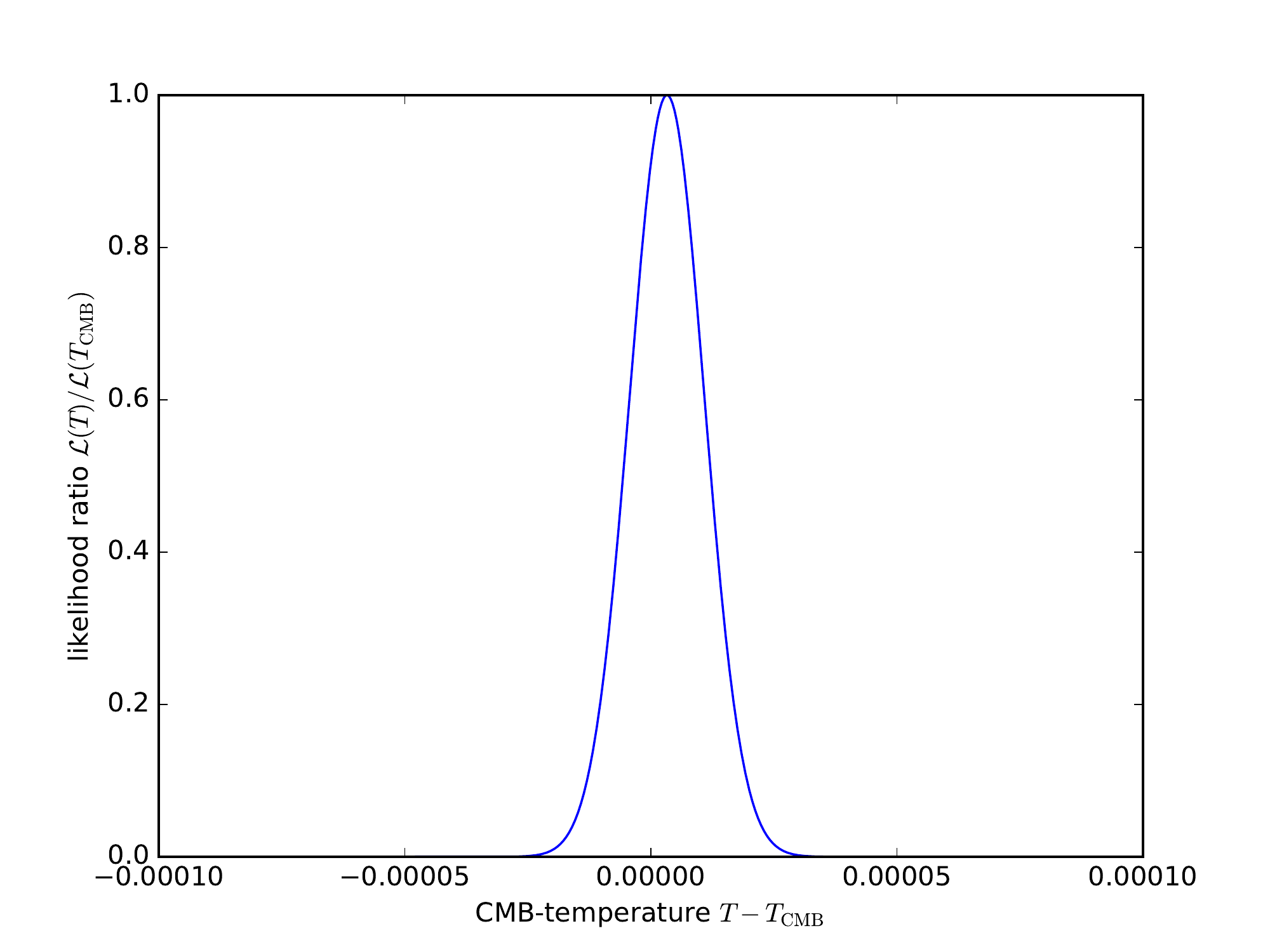}}
\caption{likelihood of the CMB-temperature $T$ for the COBE-data}
\end{center}
\end{figure}

\item \question{Solar spectrum}\\
The script \path{solar_plot.py} plots the spectrum of the Sun: Determine the surface temperature $T_\odot$ of the Sun by using Wien's displacement law and the factor that you have determined in the exercises, and estimate the error in your measurement of $T_\odot$.

\end{enumerate}

\addcontentsline{toc}{section}{Le Petit Prince}
\newdocument{Exercises}





\begin{center}
\LARGE \textbf{Astronomy from 4 Perspectives: the Dark Universe}
\HRule
\end{center}
\begin{flushright}
prepared by: Heidelberg participants and Bj{\"o}rn Malte Sch{\"a}fer
\end{flushright}
\begin{center}
{\Large \textbf{Exercise: The planet of the Petit Prince}}
\end{center}
\vspace{5mm}

\begin{enumerate}[\itshape \bfseries 1.]

\item \question{Gravity on the planet of the Petit prince}\\
The Petit Prince by A. de Saint-Exup{\'e}ry lives on a planet which,
according to images, is roughly $R\simeq 1~\mathrm{m}$ in size and
because Saint-Exup{\'e}ry does not provide any other information, has a
value of the surface gravity $g=9.81~\mathrm{m}/\mathrm{s}^2$ similar
to Earth. But in comparison to Earth where the gradient of the
acceleration is almost zero, it is much stronger on the planet of the
Petit Prince. Recall that $G=6.6\times 10^{-11}$ in SI.
\begin{enumerate}[(a)]
\item{What is the density $\rho$ and mass $M$ of the planet, assuming that it is uniform? What astrophysical objects would have similar densities?}
\item{What would be the orbital velocity $\upsilon$ of an object at a height of $1~\mathrm{m}$ above the surface? Could the Petit Prince throw an object horizontally and have it orbit his planet?}
\item{Can the Petit Prince leave the planet by jumping into space?}
\item{Is it possible that the Petit Prince can observe 43 sunsets each day despite the centrifugal force? How many sunsets can one observe at most?}
\end{enumerate}

\item \question{Devices on the planet of the Petit prince}\\
Imagine that Saint-Exup{\'e}ry brings simple mechanical systems with him, and find out if they behave differently because of the strong gradient $\partial g/\partial r$ in the gravitational acceleration $g$.
\begin{enumerate}[(a)]
\item{What's the relation between the oscillation period $T$ of a pendulum clock as a function of height $h$? Would the oscillation period be independent from the amplitude?}
\item{Saint-Exup{\'e}ry and the Petit Prince have a glass of orange juice
    with an ice cube. The Petit Prince's ice cube swims higher or not above the surface of the juice compared to Saint-Exup{\'e}ry's?}
\end{enumerate}

\item \question{Relativity on the planet of the Petit prince}\\
Are there relativistic effects of gravity on the planet of the Petit Prince?
\begin{enumerate}[(a)]
\item{What is the tidal gravitational acceleration between the head
    and the feet of the Petit Prince? Please compute the difference
\begin{equation}
\Delta g = \frac{GM}{R^2}-\frac{GM}{(R+1)^2} 
\end{equation}
}
\item{What is the gravitational time dilation between the head and the feet of the Petit Prince? Please use the formula
\begin{equation}
\Delta \tau = \sqrt{1+2\frac{\Phi}{c^2}}\:\Delta t
\end{equation}
and approximate the potential as homogeneous, $\Phi = g\Delta r$.
}
\end{enumerate}
\end{enumerate}

\newdocument{Solutions}





\begin{center}
\LARGE \textbf{Astronomy from 4 Perspectives: the Dark Universe}
\HRule
\end{center}
\begin{flushright}
prepared by: Heidelberg participants and Bj{\"o}rn Malte Sch{\"a}fer
\end{flushright}
\begin{center}
{\Large \textbf{Solutions: The planet of the Petit Prince}}
\end{center}
\vspace{5mm}

\begin{enumerate}[\itshape \bfseries 1.]

\item \question{Gravity on the planet of the Petit prince}\\
The Petit Prince by A. de Saint-Exup{\'e}ry lives on a planet which,
according to images, is roughly $R\simeq 1~\mathrm{m}$ in size and
because Saint-Exup{\'e}ry does not provide any other information, has a
value of the surface gravity $g=9.81~\mathrm{m}/\mathrm{s}^2$ similar
to Earth. But in comparison to Earth where the gradient of the
acceleration is almost zero, it is much stronger on the planet of the
Petit Prince. Recall that $G=6.6\times 10^{-11}$ in SI.
\begin{enumerate}[(a)]
\item{The relation between the mass and the density of the planet is
    $M=(4\pi /3)R^3$. The surface gravity is tied to the mass by
    $g=GM/R^2$. Substituting one in the other one can solve to find a
    density $\rho \simeq 3.5\times 10^{10}$ kg m$^{-3}$. This is quite
  close to the density of a White Dwarf.}
\item{The orbital period and velocity can be obtained by equation the
    gravitational acceleration to the centrifugal acceleration:
    $GM/(R+1\mathrm{m})^2 = \Omega^2 (R+1\mathrm{m}) $. This gives
    $\Omega \simeq 1$ s$^{-1}$ corresponding to a period $P\simeq 6$ sec, and
    an orbital speed $V=\Omega (R+1\mathrm{m})\simeq 4$ m
    s$^{-1}$. Yes the Petit Prince can throw an object fast enough to
    put it into orbital motion.}
\item{The escape speed is given by equating the specific kinetic
    energy $V^2/2$ to the potential energy $GM/R$, and this gives a
    typical value $V\simeq 4$ m
    s$^{-1}$. This is too much for a kid to jump.}
\item{Given that the maximum period is 6 sec (computed above, and our
    day corresponds to $86400$ sec then
    there will be at most $14000$ sunsets/sunrises.}
\end{enumerate}

\item \question{Devices on the planet of the Petit prince}\\
Imagine that Saint-Exup{\'e}ry brings simple mechanical systems with him, and find out if they behave differently because of the strong gradient $\partial g/\partial r$ in the gravitational acceleration $g$.
\begin{enumerate}[(a)]
\item{For small oscillations the formula for the oscillation period of
  a pendulum is $T=2\pi\sqrt{L/g}$ where $L$ is the length of the
  pendulum and $g$ is the gravitational acceleration. On this planet,
  at an heigh $h$ above the surface it will be:
  $g=GM/(R+h)^2$. Then $T=2\pi (R+h)\sqrt{L/GM}$. The period increases
  proportionally to the height. Given that $g$ is not constan with
  height, as the pendulum oscillates it will experience different
  acceleration (stronger at the bottom point, weaker at the edge points
  of its oscillating trajectory) so the pendulum formula does not
  apply, and there will be a dependence on the oscillation
  amplitude. One can try to use a mean values for $g$. Using some
  basic trigonomentry (see Figure 2) the difference in heigth between the bottom
  point and the edge point is $L(1-\cos{\theta})$, where $\theta$ is
  the amplitude of the oscillation. So the average $g \simeq
  GM/R^2[1+L(1-\cos{\theta})]/2R$. substituting this in the equation
  for the pendulum period $T$ one gets an estimate on how it depends
  on the apmplitude $\theta$.}
\item{Archimedes principle says that a body immersed in a liquid
    (water in our case) received a lift upward with a force equal to
    the weight of the volume of the liquid it displaces. With
    reference to the figure, let us consider an icecube of an edge of
    length $L$, floating in water with a depth equal to $h$ (see
    Figure 2). Let us
    call $\rho_{\rm I}$ the density of ice and $\rho_{\rm W}$ the
    density of wager. We know that ice is lighter than water, such
    that the former flots on latter. For conveneince we assume that
    the icecube is small enough that the gravity can be taken as
    uniform over its size. A typical icecube is $\sim 1$cm, while on
    the Petit Price planet the typical scale for the variation of
    gravity is $\sim 1$m. The gravitatinal force acting on
    the icecube is equal to $F_g=g\rho_{\rm I}L^3$. Archimedes force
  is instead $F_a=g\rho_{\rm W} h L^2$. The cube will float if the two
are equal, and this gives $h= L\rho_{\rm I}/\rho_{\rm W}$, that as one
can see does not depend on the graviational acceleration. So it does
not matter if the graviational field is stronger or weaker. The
icecube of the prince will float as much as the one of Saint Exupery.}
\end{enumerate}

\item \question{Relativity on the planet of the Petit Prince}\\
Are there relativistic effects of gravity on the planet of the Petit Prince?
\begin{enumerate}[(a)]
\item{What is the tidal gravitational acceleration between the head
    and the feet of the Petit Prince? Please compute the difference

\begin{equation}
\Delta g = \frac{GM}{R^2}-\frac{GM}{(R+1)^2} 
\end{equation}
}
\item{What is the gravitational time dilation between the head and the feet of the Petit Prince? Please use the formula
\begin{equation}
\Delta \tau = \sqrt{1+2\frac{\Phi}{c^2}}\:\Delta t
\end{equation}
and approximate the potential as homogeneous, $\Phi = g\Delta r$.
}
\end{enumerate}
\end{enumerate}

\end{document}